\documentclass[journal]{IEEETran}
\usepackage[pagewise,switch]{lineno}
\IEEEoverridecommandlockouts
\usepackage[noadjust]{cite}
\usepackage{amsmath,amssymb,amsfonts}
\usepackage{algorithmic}
\usepackage{graphicx}
\usepackage{textcomp}
\usepackage{xcolor}
\usepackage[capitalise]{cleveref}
\usepackage{siunitx}
\usepackage{subcaption}
\usepackage{xurl}

\usepackage{flushend}
\crefname{subsection}{Subsection}{Subsections}

\DeclareSIUnit\operations{op}
\usepackage[colorinlistoftodos,prependcaption,textsize=tiny]{todonotes}

\def\BibTeX{{\rm B\kern-.05em{\sc i\kern-.025em b}\kern-.08em
    T\kern-.1667em\lower.7ex\hbox{E}\kern-.125emX}}
\begin{document}

\title{Time-based GNSS attack detection}

\author{
	\IEEEauthorblockN{
		Marco Spanghero and Panos Papadimitratos,~\IEEEmembership{Fellow,~IEEE}
	}
	\\
	\IEEEauthorblockA{Networked Systems Security (NSS) Group -- KTH Royal Institute of Technology, Stockholm, Sweden \\
		marcosp@kth.se, papadim@kth.se}
}

\maketitle

\begin{abstract}
To safeguard Civilian Global Navigation Satellite Systems (GNSS) external information available to the platform encompassing the GNSS receiver can be used to detect attacks. Cross-checking the GNSS-provided time against alternative multiple trusted time sources can lead to attack detection aiming at controlling the GNSS receiver time. Leveraging external, network-connected secure time providers and onboard clock references, we achieve detection even under fine-grained time attacks.
We provide an extensive evaluation of our multi-layered defense against adversaries mounting attacks against the GNSS receiver along with controlling the network link. We implement adversaries spanning from simplistic spoofers to advanced ones synchronized with the GNSS constellation. We demonstrate attack detection is possible in all tested cases (sharp discontinuity, smooth take-over, and coordinated network manipulation) without changes to the structure of the GNSS receiver. Leveraging the diversity of the reference time sources, detection of take-over time push as low as \SI{150}{\micro\second} is possible. Smooth take-overs forcing variations as low as \SI{30}{\nano\second/\second} are also detected based on on-board precision oscillators.
The method (and thus the evaluation) is largely agnostic to the satellite constellation and the attacker type, making time-based data validation of GNSS information compatible with existing receivers and readily deployable.

\end{abstract}

\section{Introduction}

Global Navigation Satellite Systems (GNSS) provide ubiquitous localization, navigation, and synchronization but are still vulnerable to adversarial manipulation (\cite{Thombre2018, psiaki2016gnss}). Civilian GNSS receivers (at the time of writing) largely rely on unprotected signals (e.g. without cryptographic enhancements at the physical layer or the navigation messages \cite{Fernandez-Hernandez2016,anderson2017chips}). The Galileo Open Service Navigation Message Authentication and the upcoming GPS Chimera aim at improving this situation (\cite{Gamba2021ComputationalPlatforms,Cucchi2021AssessingReceiver,Motella2020AReceiver}), however, operational deployment requires time \cite{Gotzelmann2021GalileoProvision}. Nevertheless, cryptographic methods in newly deployed systems will not cover receivers already present in the field, when modifications to the receiver structure are needed. 
Furthermore, even when authenticated signals are considered, cryptographic protection cannot fully address replay/relay attacks (\cite{Lenhart2022,Seco-Granados2021,10.1145/3558482.3590186}).

GNSS receivers are often integrated into network-connected devices capable of various degrees of computational power: in principle, the augmented GNSS receiver can validate the GNSS-provided Position, Navigation, and Timing (PNT) by comparing it with other reference sources.
Specifically, we aim at detecting and containing attackers capable of tampering with the GNSS receiver time solution, whether by synchronized signal lift-off or time skips: if such manipulation causes the time part of the solution to drift from the correct time, protection is practical. Furthermore, the challenge is to detect sophisticated adversaries aiming at spoofing the GNSS receiver while, possibly, controlling remote time providers, in a coordinated manner. Such adversaries evade existing time-based defense mechanisms.

As known vulnerabilities can be exploited to manipulate unprotected remote time references, trustworthy reference time information and secure time transfer methods \cite{LakshayN:TH:2018} are necessary; in addition to local timekeeping, i.e., embedded oscillators. On the other hand, local time-based tests alone cannot thwart GNSS attacks at \textit{cold start}. Time information from external, networked servers, as long as those can be authenticated, can mitigate networked-based attacks combined with GNSS attacks. Nonetheless, intermittent network connectivity can deprive the receiver of such external time sources. Naturally, one can combine the two ideas, GNSS-provided time validation based on recurring interaction with networked time servers and, in the meantime, validation based on the local clock hardware.

Building upon the method in \cite{spangheroMPPPLANS23}, this work provides a comprehensive evaluation of time-based GNSS attack detection.  We consider heterogeneous clock sources studying their combination based on accuracy, security, and availability. By profiling the individual time reference providers, we optimize the attack detection threshold, the needed computational power, and energy consumption. We show a practical implementation of a testbed using existing cryptographically secure coarse time references (Roughtime \cite{ietf-ntp-roughtime-07}), online secure (or insecure) time servers (NTS \cite{ietf-ntp-using-nts-for-ntp-28} and NTP \cite{rfc5905}), and an accurate local, on-board clock ensemble to provide reliable GNSS attack detection \cite{spangheroMPPPLANS23}. 

Extending \cite{spangheroMPPPLANS23}, this work contributes the following:
\begin{itemize}
    \item An improved, refined method that now considers simultaneously all the time sources available to the system instead of a staged approach (\cref{section:methodology}).
    \item An implementation of a fine-grained GNSS attacker capable of controlling the GNSS receiver while controlling the network connectivity at the GNSS-enabled platform and we evaluate the performance of the detection system against different classes of attacks (\cref{section:sys-adversary-model}). 
    \item A complete evaluation of the improved method and system against the stronger attacker. We show how optimization of the time source selection is possible based on the quality of the remote and local time references. Additionally, we discuss the security trade-offs of the presented solution, regarding properties of remote time references (\cref{section:results-conclusion},  \cref{subsection:adaptive-sampling,subsection:security-considerations}).
    \item Overall, a thorough feasibility investigation showing the method can detect misbehaving GNSS-based time solutions under all the tested scenarios, both for sharp discontinuities of the PNT solution and sustained smooth takeover (\cref{section:results-conclusion}). 
\end{itemize}


\section{Related Work}
\label{section:background}

A brief discussion on attacks, followed by an analysis of orthogonal countermeasures that can co-exist with the method explored here prefaces the discussion of related work validating GNSS PNT with the help of local and remote time sources. 

\textbf{Attacking GNSS receivers} - Successful overtake of GNSS receivers in the field is possible by spoofing attacks that can be implemented either with signal generation or replay/meaconing (\cite{tippenhauer2011requirements, Kerns2014, Bhatti2017, Ioannides2016}). The risk of low-sophistication attacks in the wild is high given the availability of low-cost software-defined radio (SDR) hardware and software tools which are openly available (\cite{KexiongAllBelongToUs2018, Feng2021, HuangL2015}) even in the case of multi-constellation \cite{LeksellTGalileo2021} and multi-frequency modes \cite{SDRMultiFrequency2018}.

Advanced implementations of receiver-spoofer matched adversaries (\cite{HumphreysAssessingSpoofer, Maier2018}) rely on signal lift-off techniques: this requires code phase and Doppler shift synchronization at the victim antenna phase center. Additionally, deployment of attacks targeting mobile victims is complex, as precise tracking of the victim antenna is required throughout the attack. In a simpler setting, attacks targeting static timing-dedicated receivers smoothly deceive and control the time solution at the GNSS receiver, with the relevant case of phasor measurement units in smart grids (\cite{Shepard2012c, Humphreys2012, Jiang2013, Zhu2016}). Time Synchronization Attacks (TSA) target the time solution of the GNSS receiver, minimally disturbing the location or navigation part \cite{Zhang2013}. Even if centimeter-level knowledge of the victim's antenna position is required to perform a successful synchronized lift-off, code/doppler frequency sweep takeover is possible in commercial receivers. This approach eliminates the need for precise alignment of the spoofer signal and leverages the higher tracking bandwidth of modern receivers to successfully implement the attack \cite{Jiadong2019}.



Effective overtake, albeit to a lesser level of control, can be achieved also by signal replay/relay (meaconing): the attacker re-transmits signals corresponding to a different time and/or place at the victim receiver, causing a shift in the PNT solution \cite{Lenhart2022}. Such methods, in a more advanced configuration known as Secure Code Estimation and Replay (SCER) are effective also against cryptographically protected navigation messages and signals (\cite{humphreys2013detection, Arizabaleta2019, Gallardo2020}). Similarly, Distance Decreasing attacks cause significant alterations of the GNSS PNT, even against cryptographically enhanced signals(\cite{ZhangLP:J:2022, ZhangP:C:2019a}).

\textbf{Safeguarding GNSS receivers} - To counteract the growing issue of GNSS manipulation, countermeasures to achieve higher PNT solution robustness exist. Carrier-over-Noise ($C/N_0$) analysis with joint measurement of the receiver front-end gain allows detection of spoofing signals observing the power envelope variation and distortion (\cite{Akos2012, Lo2019, wesson2017gnss}). So-called Doppler shift tests in the received signals allow the detection of spoofed signals based on the transmitter frequency error (\cite{papadimMilcom2008, psiaki2013antenna}). Additionally, Receiver Autonomous Integrity Monitoring (RAIM) techniques (\cite{Jada2021, Sathaye2020}) allow detection and exclusion of adversary-crafted GNSS signal, even in case of time-specific faults \cite{gioiaTRAIM2021}. Additionally, multi-constellation failure detection \cite{ZhangP:C:2019b} and collaborative/cooperative detection methods proved capable of faulty signals detection and exclusion (FDE), hardening the GNSS receiver against adversarial manipulation. Albeit effective, RAIM and FDE methods require significant computation and access to the raw measurement from the GNSS receiver: most consumer-grade GNSS receivers do not integrate native RAIM capability or do not expose raw measurements to the user.


\textbf{Time-based GNSS PNT validation} - Correction services in the L-band contribute to the robustness and accuracy of the PNT solution \cite{rugamer2023}: L-band corrections leverage an extensive fixed receiver network to provide Real Time Kinematic (RTK) corrections over the network. Methods relying on short-range networks to dedicated receivers or internet-provided correction streams are complemented by satellite-downlink-provided corrections, whose availability is increasing even in the consumer market and are designed to provide centimeter-level accuracy for precise positioning. For time-focused corrections, recent products like Fugro Atomichron promise accurate and reliable time and frequency, but cannot be evaluated at this time. \cite{fugroAtomichron}. On the other hand, its recent introduction combined with the integration required by the receiver manufacturers will be a major limitation towards the adoption of this system. Generally, while L-band correction services provide accurate aiding information, they require dedicated hardware and modems to operate. This is a major limitation towards adoption in consumer devices and generally low-power devices that are limited. For the attacker, correction services can be effectively used to precisely know the position of the intermediate spoofer antenna, allowing accurate estimation of the lever arm (vector between the reference and victim antenna). This leads to a more precise estimation of the victim's position, aiding the overtake of the victim receiver.

Commonly available connectivity (i.e., via cellular network) can be used to access alternative PNT information securely, which intuitively can be leveraged for GNSS receiver-provided time validation. Solutions considering single or multiple precision embedded clocks proved successful in detecting offsets and drift in the time solution due to an adversary (\cite{Arafin2016DetectingOscillators, Arafin2017, Spanghero2022, Hwang2014}). Receiver autonomous testing of the GNSS clock bias and drift allows monitoring of abrupt changes in the receiver clock bias \cite{jafarnia2013PNT}. While this is often an indicator of spoofing or other adversarial action, it requires the receiver front end to be disciplined with a high-quality local oscillator and rely on moving antennas. While the second is common in mobile devices but is not applicable for fixed installations, the first is usually difficult to achieve in commercial receivers, given that there is no access to the clock interface.

Network time providers enable even sparsely connected receivers to test the accuracy of the GNSS-provided time in respect to a set of remote references (\cite{spangheroGNSS20,kzmsppPLANS2020, spangheroMPPPLANS23}). Such countermeasure complements and augments other methods based on signal properties (\cite{papadimMilcom2008}) and can be integrated into existing GNSS-enabled platforms without changes to the receiver structure or the existing hardware.  Performance assessment of secure time transfer in support of cryptographically enhanced GNSS signals shows that the accuracy of network-provided time is sufficient for use with Chimera \cite{ODriscol2020}. Furthermore, the application of a combination of diverse time references to the current GNSS receivers and signals proved capable of hardening the security of the receiver \cite{spangheroMPPPLANS23,patentPP:J:2009}, but a systematic analysis of the actual capabilities against adversaries targeting both the GNSS receiver and alternative time providers is missing. This work sets out to validate in an extended experimental context \cref{section:implementation} the methodology shown in \cite{spangheroMPPPLANS23,kzmsppPLANS2020} when not limited to a staged approach but by considering as a whole any time reference available to the system (as shown in \cref{section:methodology}). An extended advanced attacker model (\cref{section:sys-adversary-model}) is used in a comprehensive evaluation of the results (\cref{section:results-conclusion} to demonstrate that time-based cross-check of the GNSS PNT solution against heterogeneous time sources is practical.



\section{System and Adversary model}
\label{section:sys-adversary-model}
The system model recalls the setup from \cite{spangheroMPPPLANS23} by considering an off-the-shelf consumer GNSS receiver connected to a computation and network connectivity platform. We assume the system has one or more onboard precision reference oscillators to provide high-quality, stable reference time. When not under adversarial control, the GNSS receiver is always the most accurate PNT source available to the system. We assume the system always uses the most accurate PNT available unless a discrepancy against any reference source is detected and the PNT is deemed to be under attack. Under such conditions, the most trusted time reference is selected.

As the time sources available to the system (which can consist of either enhancements to the system components, or provided by network-connected entities, beyond the GNSS receiver) are different, clear assumptions are needed for 
the level of trustworthiness and performance required. Time references that are within the hardware boundaries of the GNSS-enabled system (e.g. its packaging) are trusted, meaning the hardware device cannot be compromised. Time sources external to the GNSS-enabled system (e.g., network-provided time) are deemed not trusted unless cryptographically protected, authenticating the communication and validity of the time information. 
Still, the system can use both trusted and untrusted network time providers, but in case of a discrepancy among the time solutions, it will resort to the most trusted provider, even at a penalty of reduced accuracy. Furthermore, we do not make any strict assumption on connectivity (which could be unavailable due to a benign fault or adversarial action), hence we do not require a constant exchange of information with the connected time references. 

Additionally, at startup, the receiver can have no current knowledge of the state of the constellation and the time offset of its embedded oscillator (cold start) or the receiver has already acquired a valid solution and obtained recent constellation status updates. 

We consider the case of a single constellation GNSS receiver, specifically GPS: this does not limit the scope of the countermeasure shown here as it is compatible with different GNSS systems or even combinations of multiple constellations (i.e. multi-constellation GPS and Galileo receivers). 
The objective of the receiver is to solve \cref{eq:PNT_eqn_good}, where $p$ is a $n \mathrm{x} 1$ vector of pseudoranges observations, \textcolor{black}{$H$ is an $n\mathrm{x}4$ observation matrix}, $x=[x,y,z,t]$ is the receiver state vector of location and time offset at the victim receiver, and $v$ is the total system noise.

\begin{equation}
    p = Hx + v
    \label{eq:PNT_eqn_good}
\end{equation}

Precise timekeeping is key for all GNSS systems to measure the distance between the receiver and each satellite. In \cref{eq:pseudorange}, the satellite-receiver pseudorange for satellite (s) in the $i$-th band is a function of the receiver's time at signal reception time, $\bar{t}_r$.
\begin{equation}
    P^{(s)}_{r,i} = c(\bar{t}_r - \bar{t}^{(s)})
    \label{eq:pseudorange}
\end{equation}

\cref{eq:pseudorange} can be expressed as a function of the geometrical range $\rho_r^{(s)}$, the receiver and satellite clock biases $dt$ and $dT$ respectively (for simplicity, we purposely exclude any atmospheric correction terms as they only represent additive errors). The resulting equation is shown in \cref{eq:pseudorange_geometerical}, where $\epsilon_p$ represents all the measurement and instrumentation delays in the system.

\begin{equation}
    P^{(s)}_{r,i} = \rho_r^{(s)} + c(dt_r(t_r) + dT^{(s)}(t^{(s)})) + \epsilon_p
    \label{eq:pseudorange_geometerical}
\end{equation}

The transmission time (\cref{eq:pseudorange_geometerical}) at the satellite can be obtained based on the Time of Week (TOW) in the navigation message, once the tracking loops lock in the satellite signals, while the receiver time offset (e.g., the time difference between the receiver's time and the satellite time) is progressively refined. Based on the information obtained from the Locked Delay Loop and its reference clock, the GNSS receiver can obtain the actual pseudorange for one reference satellite. Based on this reference signal, the GNSS receiver derives the relative reception time of all the other satellites and their pseudoranges. Finally, it can calculate the full PNT solution. Once the receiver is fully locked to the GNSS signal, and has a PNT solution the local clock error is corrected.

An adversary, crafting GNSS signals consistent with its objective can control the GNSS receiver time. Due to the open structure of GNSS signals, an attacker can create signals with valid modulation, information content, and spectral components. Additionally, navigation message information is also open and can be obtained either on the spot by direct download from the GNSS constellation or using an internet-based reference information provider. Under spoofing conditions, the output of the correlator of the victim receiver can be modeled, for the in-phase and in-quadrature components as \cref{eq:i_spoofed_superpose,eq:q_spoofed_superpose}, where $P$ denotes the power of the receiver signal; $R$ and NAV are the autocorrelation function and the navigation message, respectively; $t,f$ are the code and frequency offset deviations; $\psi$ is the frequency tracking error and $T_{coh}$ is the coherent integration time. 

\begin{multline}
    I_v = \sqrt{P_a}\text{NAV}_aR(t_{a})sinc(f_{a}T_{coh})cos(\psi_a) + \\ \sqrt{P_s}\text{NAV}_sR(t_{s})sinc(f_{s}T_{coh})cos(\psi_s) + \eta_I
    \label{eq:i_spoofed_superpose}
\end{multline}
\begin{multline}
    Q_v = \sqrt{P_a}\text{NAV}_aR(t_{a})sinc(f_{a}T_{coh})sin(\psi_a) + \\ \sqrt{P_s}\text{NAV}_sR(t_{s})sinc(f_{s}T_{coh})sin(\psi_s) + \eta_Q
    \label{eq:q_spoofed_superpose}
\end{multline}

\textcolor{black}{ In \cref{eq:i_spoofed_superpose,eq:q_spoofed_superpose}, the notations $s$ and $a$ denote the spoofed and authentic signals respectively.}

\textcolor{black}{The modified model for the PNT solution at the GNSS receiver under adversarial control, based on \cref{eq:PNT_eqn_good}, takes into account $\mathcal{S}$, the result of the spoofed signals tracked by the receiver as shown in \cref{eq:PNT_eqn_atk}.}

\begin{equation}
    p = Hx + \mathcal{S} + v
    \label{eq:PNT_eqn_atk}
\end{equation}

For an adversary to be considered as such it needs to cause a tangible deviation in the legitimate time information either on the GNSS side or by controlling the network-provided time. This means that an adversary simply replacing legitimate signals with matching fake signals cannot be considered harmful, as the resulting time information would still be correct. Adversaries below the application requirement level (e.g. causing a deviation that can be accounted for in the normal noise level tolerated by the specific application) are often undetectable by application and data layer methods. 

An unrefined attacker would generate signals matching an $x$ solution matching the attacker's intent without any awareness of the legitimate signals. Given enough power advantage or with a combination of a short jamming pulse, the adversary first causes a loss of lock in the victim, following re-acquisition on the higher-power adversarial signals; the GNSS receiver is tricked into obtaining a fake PNT solution. Due to the lack of synchronization between the GNSS frames in the real and fake signals, sharp discontinuities in the PNT solution can be observed in the time component.

To successfully mount a synchronous attack, the attacker aligns the simulated signals with the ones that are currently being tracked by the victim receiver. To do so, two requirements need to be satisfied: code phase and Doppler shift offsets of the simulated signals need to be within the bandwidth of the tracking loop and the transmission time of the simulated signals need to match the start of the GNSS subframe. The parameters for the initialization of the adversarial signals in \cref{eq:i_spoofed_superpose,eq:q_spoofed_superpose} are obtained by an intermediate receiver: Doppler shift and current code phase at the victim are corrected to the intermediate receiver antenna position by calculating the lever arm between the estimated position of the victim receiver and the attacker antenna. Additionally, navigation bits are obtained from publicly available distribution services (e.g., NASA CDDIS, Novatel) or through the intermediate receiver itself. Given the long repetition period (\SI{12.5}{\minute}) the navigation data bits are fully predictable.


We are agnostic in terms of the specific type of attack mounted by the adversary to control the victim receiver with the following limitation: the attacker spoofs civilian (encryption-less) signals without cryptographical enhancements, which are compatible with the majority of the available receivers\footnotemark{}. \footnotetext{Other complex attacks that are possible against cryptographically enhanced receivers (such as Secure Code Estimation and Replay (SCER)) are beyond the scope of this work. An attacker can not maintain consistency of both the time and position solution at the onset of a SCER attack (\cite{psiaki2013}), making it a good candidate for detection based on consistency of the time solution.
On the other hand, the sophistication for such attacks to be successful makes them unlikely against low-value targets. Nevertheless, considering SCER attacks is worth a dedicated investigation.}
The GNSS attacker model in this work is generic and analyzes the effect of an adversary causing variations, sharp or smooth, in the receiver time solution without limiting the adversarial action's aim which could be the receiver's time itself or its position.
Any attack capable of independent modification of the position solution without disturbance of the time solution, if achieved, would be undetectable by any countermeasure monitoring the receiver clock solution, including the one discussed here\footnotemark{}. \footnotetext{This is an intrinsic limitation of any time-based countermeasure. Beyond specific attacks designed for particular conditions (e.g., static receiver, controlled environment), attacks to be expected in the wild are likely to be less sophisticated, likely causing some level of variation in the clock solution. Because of this, even if all-around assurance of the PNT solution is hard to guarantee based on the clock solution only, it is still a good indicator of likely misbehavior. Additionally, there is no limitation to combining this method with other detection schemes targeting different PNT aspects.} 

\begin{figure}[h!]
    \centering
    \includegraphics[width=\linewidth]{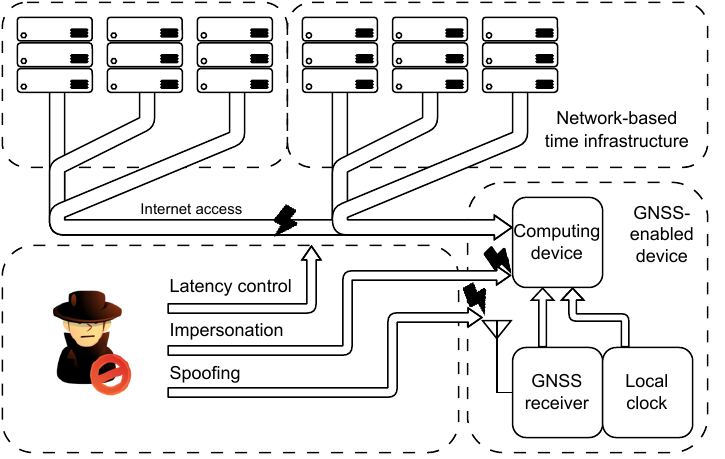}
    \caption{An adversary capable of targeting the GNSS receiver and the network-based time reference.}
    \label{fig:atk-pictorial}
\end{figure}

Additionally, a gamut of more traditional attacks targeting the synchronization and timing infrastructure exists (\cite{perry2021, Deutsch2018, Malhotra2017}), specifically focusing on network time manipulation. For this reason, we extend the attacker action beyond the GNSS receiver to the communication link between the GNSS-enabled platform and the Internet. Specifically, attacks focusing on the protocol and latency aspects of network-based time provision can be combined with time-aligned GNSS overtake to minimize the detection probability. Such an adversary can control the victim's access to the network-provided time by denying, limiting, or tampering with the network access. In addition, the attacker can impersonate one or more selected servers among the ones available to the victim. Attacks on the Network Time Protocol (NTP) show that a strategically placed attacker can modify the perceived time at the victim by fully or partially controlling the victim's interaction with the remote NTP time server (or server pool) specifically when no cryptographic protection is provided. As a result of the combination of the attack surfaces, on the GNSS receiver and the network side, a complex adversary is considered, capable of different coordinated action on multiple components of the system, as shown in \cref{fig:atk-pictorial}.

\section{Methodology}
\label{section:methodology}

Building upon \cite{spangheroMPPPLANS23}, we introduce a significant change in the way the fusion of multiple sources is performed. Intuitively, although the method in \cite{spangheroMPPPLANS23} is effective, there is no limitation in the specific order external and internal time references are combined, based on the available resources and connectivity.


The GNSS receiver continuously provides the system with PNT updates, as long as enough satellites are in view. Similarly, provided that sufficient connectivity is available, the device can track an arbitrary number of remote time references while monitoring the local oscillator. Based on this, the GNSS-enabled platform is provided at any given moment with a variety of different time providers, beyond the GNSS receiver itself, that can be used to validate the GNSS-based time. Intuitively, the system can combine whatever time reference is available, and based on the comparison with the GNSS receiver time, decide if the GNSS-based timestamp is to be deemed trusted. 

As an adversary could capture the receiver before the first legitimate fix is obtained, it is important to protect the initial acquisition process, bounding the initial PNT solution (and by extension the initial receiver offset) within an error of a trusted time source. 

\begin{figure}[h!]
    \centering
    \includegraphics[width=\linewidth]{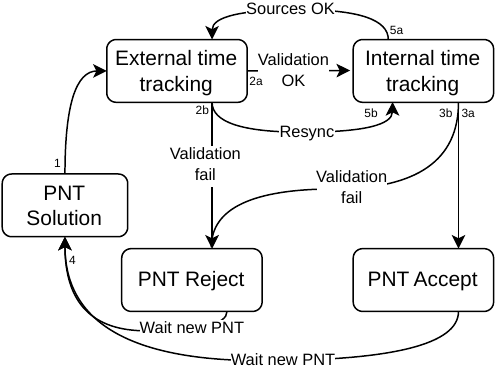}
    \caption{Logical test progression }
    \label{fig:switching-schema}
\end{figure}

The logical progression is shown in \cref{fig:switching-schema}: at startup, the platform starts monitoring the GNSS receiver and any available time reference based on its local oscillator. Once a PNT solution is available (Step 1), the device needs to test the validity of the GNSS-based time information. One immediate issue arises if no previous knowledge about the current correct time is available and no remote time reference is reachable: in such a situation, the device could already be under attack but unable to validate the absolute time offset. In such a scenario, the device can only detect changes that happen by leaving the spoofer-affected area or when the spoofing signal transmission stops, allowing the receiver to lock on real signals.

If connectivity is available (even for short time), the device tests if the GNSS-provided time is comparable with the one obtained from the available remote time source. If the check corroborates the validity of the GNSS-provided time (Step 2a), the GNSS time is tested against the local clock. Otherwise, the GNSS platform rejects the GNSS-provided time and marks the time solution as untrusted (Step 2b), until a new PNT solution (Step 4). The confidence the platform has in this decision depends on the accuracy of the available time sources and their perceived level of trust. 

At this point, the system has current knowledge of the time offset and can rely on the onboard clock to track the status of the GNSS-based time, monitoring changes in the drift or time offset. Upon successful integration of the clock parameters (Step 3a), the PNT solution is considered valid; it is otherwise discarded (Step 3b). Notably, the PNT solution can still be discarded based on the local clock information even if the validation based on the external time tracking is successful. Additionally, the system can rely on the local oscillator to keep track of drift and jitter in the reference sources (Step 5a). Periodic re-synchronization with the remote time server, based on its assumed performance and the available connectivity, allows for re-calibration of the local oscillator (Step 5b).

\subsection{External time reference providers}
\label{sec:roughtime_comp}
In regards to external time reference providers, we limit our scope to two types of time providers (Roughtime and NTP/NTS). Google Roughtime provides secure and digitally signed time information from an ecosystem of trusted providers. Notably, this is not a time transfer method (there is no two-way delay compensation) but more a secure time validation. NTS (and to a certain extent, NTP) allows secure synchronization with precise two-way delay estimation against a remote time source but requires stable connectivity.

Once the receiver solves \cref{eq:PNT_eqn_good} for $dt_r$, it obtains the time offset between the receiver clock and the GPS timescale. The PNT solution is often expressed as aligned to the UTC scale, which means that the receiver also compensates for leap seconds that correspond to the offset between the GPS timescale and UTC. Google Roughtime achieves a high level of time distribution assurance by providing digitally signed coarse time information.  To obtain secure time verification, a client creates a request to a remote time reference and obtains a digitally signed reply containing the authenticated time. A Roughtime measurement is $T^{(s)}_{RT} = \{t^{(s)}, R^{(s)}\}$, which are the absolute timestamp of UTC and the server's confidence radius, which indicates the server's estimate of its accuracy. Consider the following $t_{GNSS}, t^{(s)}_{RT}$, the receiver provided UTC, and the Roughtime timestamp. We can write the following binary hypothesis test for any Roughtime server available in the ecosystem:

\begin{equation}
    \mathcal{H}_i =
    \begin{cases}
        \mathcal{H}_0\ \mathrm{if}\ |t_{GNSS} - t^{(s)}_{RT}| < R^{(s)} \\
        \mathcal{H}_1\ \mathrm{otherwise}
    \end{cases}
    \label{eq:hypothesis-rt}
\end{equation}

The result of \cref{eq:hypothesis-rt} determines if the GNSS-provided UTC scale is aligned with a trusted UTC scale (i.e. after the initial validation, but at any epoch in time the receiver is providing a PNT solution). Even though Roughtime is not nearly as accurate as GNSS, it has the advantage, as corroborated in \cref{section:results-conclusion}, that even in adverse network conditions it can provide time reliable time estimates.
\label{sec:nts_comp}

Based on the available connectivity quality, the NTS server pools allow direct verification of the GNSS time offset.  This is performed by simply differencing the GNSS-obtained time and the one (or multiple) obtained from the NTS pool, with an approach similar to \cref{eq:hypothesis-rt}. In this case, the test, valid for each NTS server reachable at any epoch, is shown in \cref{eq:hypothesis-nts}, where $t^{(s)}_{NTS}$ is the NTS derived time and $\lambda_{T_R}$ is a threshold obtained based on the quality of the remote NTS source (which can be configured based on the application requirements).

\begin{equation}
    \mathcal{H}_i =
    \begin{cases}
        \mathcal{H}_0\ \mathrm{if}\ |t_{GNSS} - t^{(s)}_{NTS}| < \lambda_{T_R}\\
        \mathcal{H}_1\ \mathrm{otherwise}
    \end{cases}
    \label{eq:hypothesis-nts}
\end{equation}

Generally, the hypothesis is tested per each server the NTP/NTS client process on the GNSS device is monitoring, which can create issues regarding the agreement of different time sources. Regarding cross-checking of the multiple NTS/NTP servers, this can be achieved either by evaluating the output of the single hypothesis per each test or by combining the NTP/NTS time estimated as shown in \cref{section:marzullo} and then testing the hypothesis on the compound time estimate.

\subsection{Internal time reference providers}
\label{sec:local_comp}

Precise local oscillators can continuously oversee the state of the receiver. 
Their usage is two-fold: the embedded oscillator establishes a local timescale with a known offset to the GNSS time (potentially zero) and allows monitoring of the time solution even in the absence of connectivity. Second, if connectivity is present, the local oscillator can be used jointly with the external time validation and to monitor any external time reference. 
Intuitively, within the provided stability window of the local oscillator, the progression of time in the GNSS receiver and in the local timescale is identical, allowing the platform to monitor for any unexpected behavior of the GNSS-based time solution. During an attack causing the manipulation of the GNSS receiver time offset, a discrepancy is measured between the local timescale and the GNSS-provided one even without internet connectivity.
Consider the model from \cref{eq:clock_model_gnss}, normally used for a GNSS clock, where $d,b$ are the clock drift and bias respectively, and $w_d, w_b$ are the process noise values for clock drift and bias.
\begin{equation}
    \begin{pmatrix}
        \dot b \\ \dot d
    \end{pmatrix}
    =
    \begin{pmatrix}
        0 & 1 \\ 0 & 0
    \end{pmatrix}
    \begin{pmatrix}
        b \\ d
    \end{pmatrix}
    +
    \begin{pmatrix}
        w_b \\ w_d
    \end{pmatrix}
    \label{eq:clock_model_gnss}
\end{equation}

The inter-scale clock bias $\Delta_{1,2}$ between the local oscillator timescale and the GNSS one is an uncorrelated random-walk variable. A Kalman filter tracks the state of the timescale difference jointly with the frequency difference between the GNSS disciplined clock and the reference oscillator.
Over short periods, both legitimate GNSS time and onboard precision reference are stable and with negligible drift. At the update step, the system can reject new measurements based on the confidence interval derived from the covariance update step. New measurements can be discarded if $\hat{x}$ (the measured vector of offset and drift for each clock in the system) is not within the $S$ confidence values of the predicted $x$, shown in \cref{eq:equation_test_single}, where $S,H,P,R$ are the covariance, measurements, prediction, and measurement noise matrices.

\begin{equation}
    \begin{aligned}
         & S(t_{n+1}) = H(t_{n+1})P(t_{n+1|n})H^T(t_{n+1}) + R \\
         & \hat{x} = x(t_{n+1|n}) \pm diag(S(t_{n+1}))
    \end{aligned}
    \label{eq:equation_test_single}
\end{equation}

Innovation testing based on \cref{eq:equation_test_single} only considers one measurement at a time: the intuitive extension is to consider a sequence of $m$ state estimates $x$ on which a windowed statistic, and estimate variance and expected for all values within the window. This proves to be a powerful yet computationally expensive tool \cite{spangheroMPPPLANS23}, which helps in the case of lower-quality local oscillators.

\subsection{Extracting remote clock properties using Allan variance}

\label{sec:combining}
\begin{figure}
    \centering 
    \includegraphics[width=0.9\columnwidth]{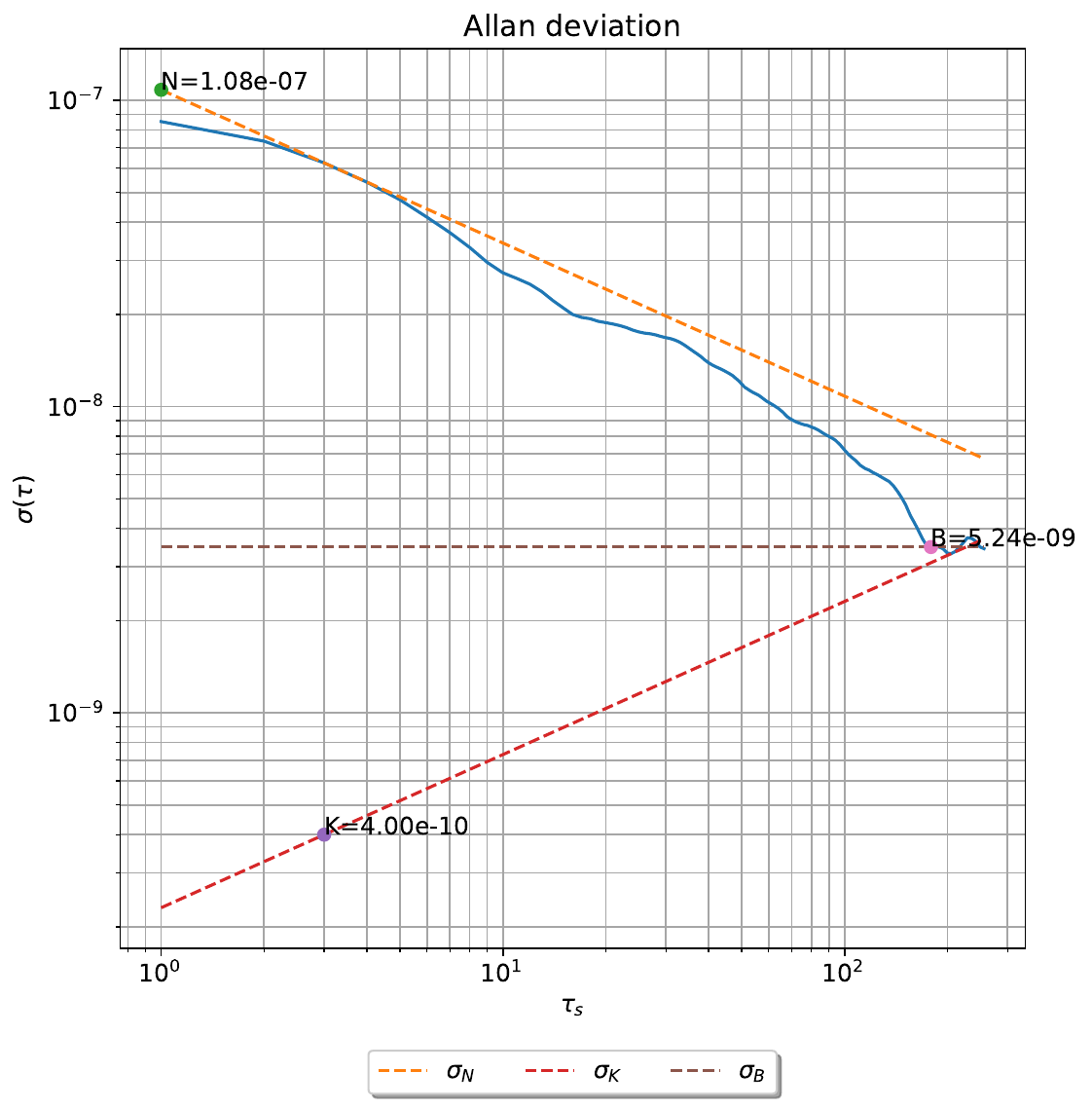}
    \caption{Allan deviation based on \cref{eq:allanvar} for reference PPS source, intercept points at noise transitions.}
    \label{fig:adev-ref-pps-unspoofed}
\end{figure}

From a practical standpoint, periodic checking of the GNSS time solution against a set of trusted time sources can be done by simple comparison. Several issues arise in doing so, which we highlight here. First and foremost, it is challenging to meaningfully compare time sources that declare different levels of accuracy and stability. As an example, a healthy GNSS-provided time pulse has a resolution in the order of \SI{10e-6}{\second} (\cref{fig:adev-ref-pps-unspoofed}), while an internet-provided NTP server can provide only \SI{10e-5}{\second} (at best) accuracy, depending on the network path latency. Second, not all time sources provide time in the same way: relative frequency stability and absolute time stamping are not necessarily ubiquitous but depend on the specific time source and possibly the network condition for external time references.

One major difficulty in establishing "clusters" for different time sources comes from the observability of the individual clock parameters: testing an oscillator's quality and stability requires observing the behavior of the clock over long periods. Time and frequency quality characterization is often performed using a clock whose intrinsic properties are better than the one under test. For this reason, the extraction of quality measurements based on sparse observations where the local time reference is often of lower quality than the remote one (which can either be the GNSS system or the network time infrastructure) is difficult and prone to artifacts caused by the local clock. 

Practically one needs a good indicator of the quality of the remote time source to be used as a binning metric to establish a quality ranking between different time sources, allowing progressive refinement of the belief in a specific time reference based on the length of the observation period.

The Allan variance is a statistic metric commonly used in the frequency domain to easily interpret the quality of a clock source \cite{Allan1987}. Specifically, it is adimensional, it provides information regarding the measurement interval and allows for direct comparison of heterogeneous clock sources, capturing not only the stability but the noise sources and their type and can be used to rank the available time sources based on their quality over time. Allan variance is defined as \cref{eq:allanvar}, where $\tau$ indicates the integration step (usually defined in decades) and $\langle...\rangle$ denotes the expectation operator.

\begin{equation}
    \sigma^2_y(\tau) = \frac{1}{2\tau^2}\langle(x_{n+2} - 2x_{n+1} + x_n)^2\rangle
    \label{eq:allanvar}
\end{equation}

It is important to note how meaningful comparisons can be extracted between clocks only where they exhibit the same behavior, defined by the accuracy over the integration period. 
\textcolor{black}{This can be obtained by direct analysis of the Allan deviation, notably its slope. There are characteristic values at which it is evaluated: the White Noise rate ($\sigma_N$), where the Allan deviation slope of $-0.5$, Flicker Noise ($\sigma_K$), with slope of $0.0$ and Random Walk ($\sigma_B$), with Allan deviation slope of $0.5$}. 
\textcolor{black}{Finally, the integration period beyond which the clock accuracy is dominated by drift is characterized by an Allan deviation slope of $+1$. In this region, the clock error grows too fast to be a used as a valid time reference against the GNSS time solution. }
\textcolor{black}{Such indicators give visual and immediate feedback on the quality of a time source or its behavior when under adversarial control. Further detail on the derivation, calculation and evaluation of such parameters can be found in \cite{804271}.} 

Two clocks exhibiting comparable values of the Allan deviation at the same integration time are likely to exhibit the same properties. Consequentially, quality indicators extracted from the Allan deviation can be used to cluster different clocks at different integration times. Intuitively this means that the system tries to cluster together clocks that are comparable at the specific observation interval the validation system is operating (e.g. we compare every \SI{60}{\second} clocks that exhibit similar drive and frequency offset when evaluated at the same interval). This is specifically true for external time sources, whose quality depends on their clock accuracy, network access, and congestion. 

\subsection{Solution testing based on multiple references}

\begin{figure}[h]
    \centering 
    \includegraphics[width=\columnwidth]{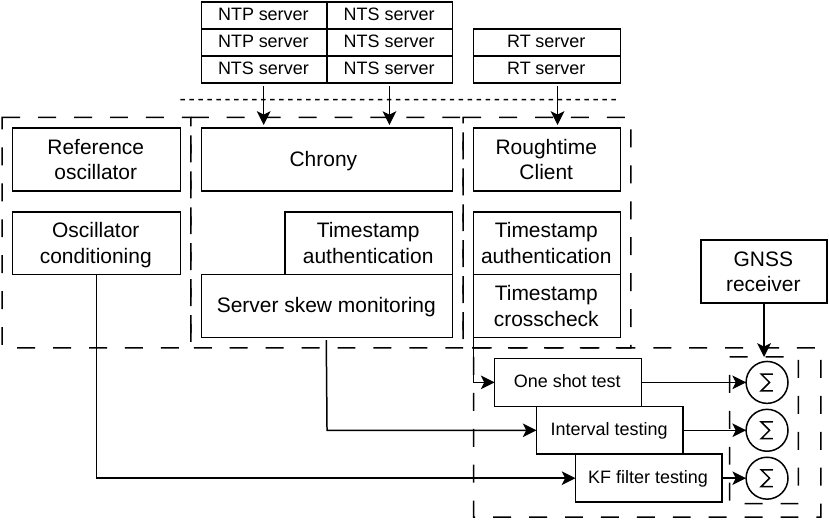}
    \caption{Solution testing based on multiple references.}
    \label{fig:fusion-schema}
\end{figure}

\cref{fig:fusion-schema} shows how the combination of multiple sources is achieved. Each time reference is authenticated (if provided with cryptographic information) and independently monitored to keep track of the individual clock performances allowing for minimal self-check on the clock source. The latter is achieved by monitoring the skew and by continuously updating the Allan deviation metric. Each timestamp, based on its properties, is fed to the test block that compares it with the GNSS-provided time solution (based on the approach devised in \cref{section:methodology}) providing the final application with a security and robustness estimation of the GNSS receiver's timing. 
\label{section:fusion-tradeoff}

\label{section:marzullo}
Once quality indicators are obtained using the Allan deviation, it is possible to compare the PNT solution provided by the GNSS receiver against any available time provider. While the local clock has known and stable properties and its availability is guaranteed, this is not necessarily true for the external time references which can be added and removed to the time check. Similarly to what is adopted in NTP clock selection, Marzullo's algorithm \cite{rfc5905} can be exploited to decide if there is an agreement between the multiple network-based references and the GNSS receiver's time.

The selection of overlapping time sources is based on two parameters $\theta, \rho_d$, the declared time, and the root distance respectively (for consistency with the IETF NTPv4 standard, as defined in \cite{rfc5905,ntp_perf}). For each available time source, we calculate the upper and lower bounds of the declared time interval as $[\theta - \rho_d, \theta + \rho_d]$. Differently from the Marzullo algorithm, we take as a reference point the GNSS-provided time and consider how many overlapping intervals contain the GNSS-provided one. Three outcomes exist: all intervals are in agreement with the GNSS time, only a subset of intervals contain the GNSS time, and no interval overlaps with the GNSS-provided time. In the first case, the outcome is trivial, as all servers agree with the GNSS-provided solution which is correct. In the last case, as there is an agreement on the time references, the GNSS-provided time can be considered as the misbehaving one. The remaining case requires careful consideration, based on the number of agreeing sources. This approach is suitable even in the presence of non-authenticated sources, given that it takes into account both the stability of the time source (measured using the Allan variance) and the properties of the network link, monitored by updating the root distance.


The system aims to quantify the quality of the "fused time" and the assurance level it can provide to the GNSS-provided time. While accuracy and stability are easily quantifiable the same quality metrics used in the qualification of the time sources applied to the overlapping intervals, the latter is more complex. Primarily, it requires establishing the trust level the user has for each time reference: intuitively, reference sources that rely on cryptographic methods to protect the time transport provide a higher level of trust than those that do not. The trade-off that can be set in the time fusion state machine depends on the specific requirements of the application leveraging GNSS-derived time. Based on these assumptions, the system ranks network-based time based on the level of trust and calculates the assurance level of the detector output, where assurance levels indicate the level of trust the system has in the decision of the detector based on the trust in the reference time.

It is important to note that the current consumer devices focus on availability, meaning that a time solution (e.g. GNSS-based) is always provided to the user. Critical time systems instead shift focus towards constant accuracy, meaning that the time solution provided to the user supposedly does not degrade over time. Systems that do not require maximum accuracy or availability can instead pursue a different path: the selection of the reference clock is based on the perceived level of trust a specific reference has when compared to its accuracy. This allows the user to trade-off accuracy or availability depending on the level of assurance the application requires for the time information.




\section{Implementation}
\label{section:implementation}
		
\begin{figure}[h]
    \centering 
    \includegraphics[width=\linewidth]{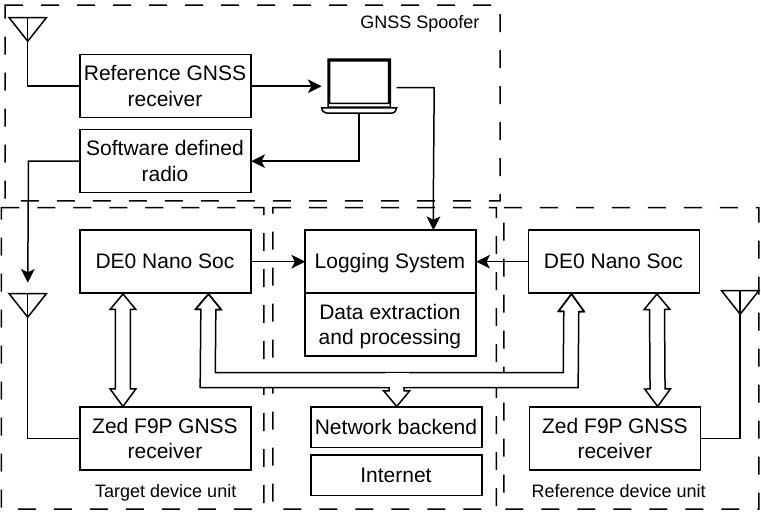}
    \caption{Experimental testbed for time-based GNSS validation.}
    \label{fig:testbed-schema}
\end{figure}

The components described in \cref{section:methodology} are implemented in an experimental testbed, as shown in \cref{fig:testbed-schema}. Additionally, the testbed implements an extended attacker based on the model provided in \cref{section:sys-adversary-model} capable of recreating advanced signal lift-off and simulation attacks jointly with disturbances on the network link of the victim.

The reference device unit in \cref{fig:testbed-schema} is used as a reference system to validate the results, while the target device unit in \cref{fig:testbed-schema} is the experiment target, connected to the GNSS spoofer.  Both systems are based on an Altera DE0-SOC FPGA and are provided with a ZED-F9P GNSS receiver. Each node has internet connectivity and the same set of remote time servers, part of a publicly available network time infrastructure. For monitoring and logging purposes, the target device unit can measure its local clock offset against the reference device unit directly, on a local network connection. 

\begin{figure}
    \centering
    \includegraphics[width=\linewidth]{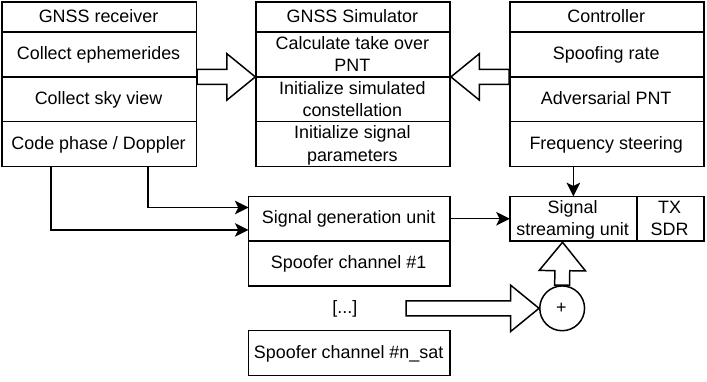}
    \caption{Spoofing system with live sky synchronization.}
    \label{fig:attacker-schematic}
\end{figure}

An attacker capable of spoofing GPS signals and controlling the network link is connected to the target device unit, which receives a combination of original and spoofed signals. The spoofer is capable of deploying both coarse spoofing signals (without any synchronization to the real signals) or refined spoofing signals (aligned with the real constellation, using the reference receiver). Due to regulatory limitations, the transmission of the spoofing signals is conducted over cable, and the attacker-simulated signals are combined with the original signals at the test device unit GNSS receiver antenna. For the refined attack, we adopt a similar strategy to \cite{Jiadong2019}, consisting of generating signals that are synchronized to the legitimate ones within the error margin of the victim receiver tracking loops. The functional blocks of the spoofer are shown in \cref{fig:attacker-schematic}. In addition, to live sky information obtained with the intermediate receiver, the radio front end and DSP clocks are synchronized, using a GPS-disciplined clock, with the real GNSS signal to lock the transmission time start with the GNSS frame start. This has a dual purpose: in addition to precise absolute transmission start time, it allows accurate frequency control of the transmitter, making it harder for the receiver to apply countermeasures based on Doppler monitoring or other signal parameters.

\begin{figure}
    \centering
    \includegraphics[width=\columnwidth]{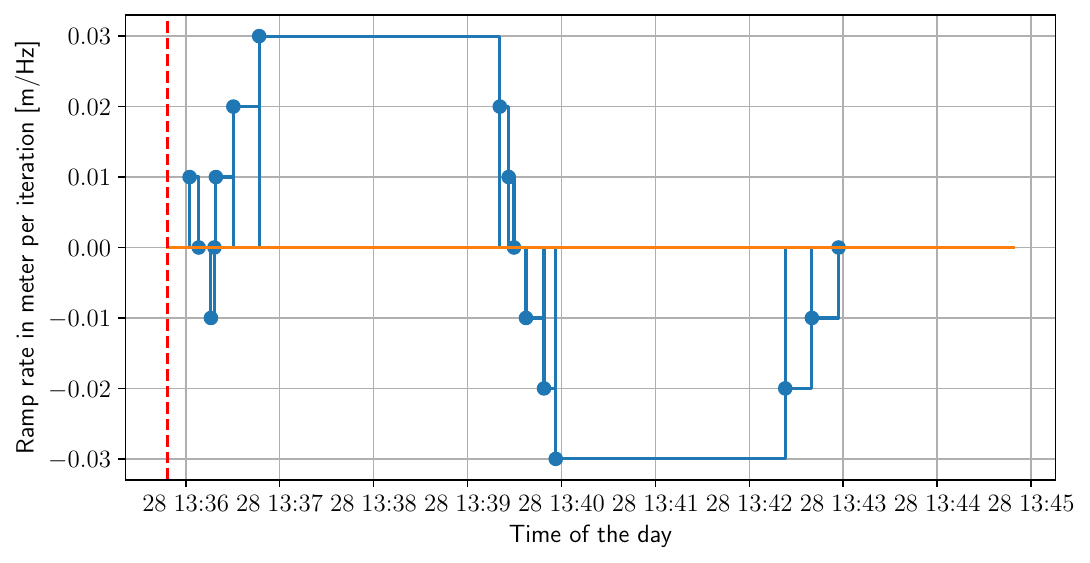}
    \caption{Example of ramp generation for pseudoranges steering: the ramp controls the rate of change of the pseudorange in meters per iteration of the spoofing signal update.}
    \label{fig:attacker-pr-strategy}
\end{figure}

\begin{figure*}
    \centering
\begin{subfigure}[b]{\columnwidth}
    \centering
    \includegraphics[width=\linewidth]{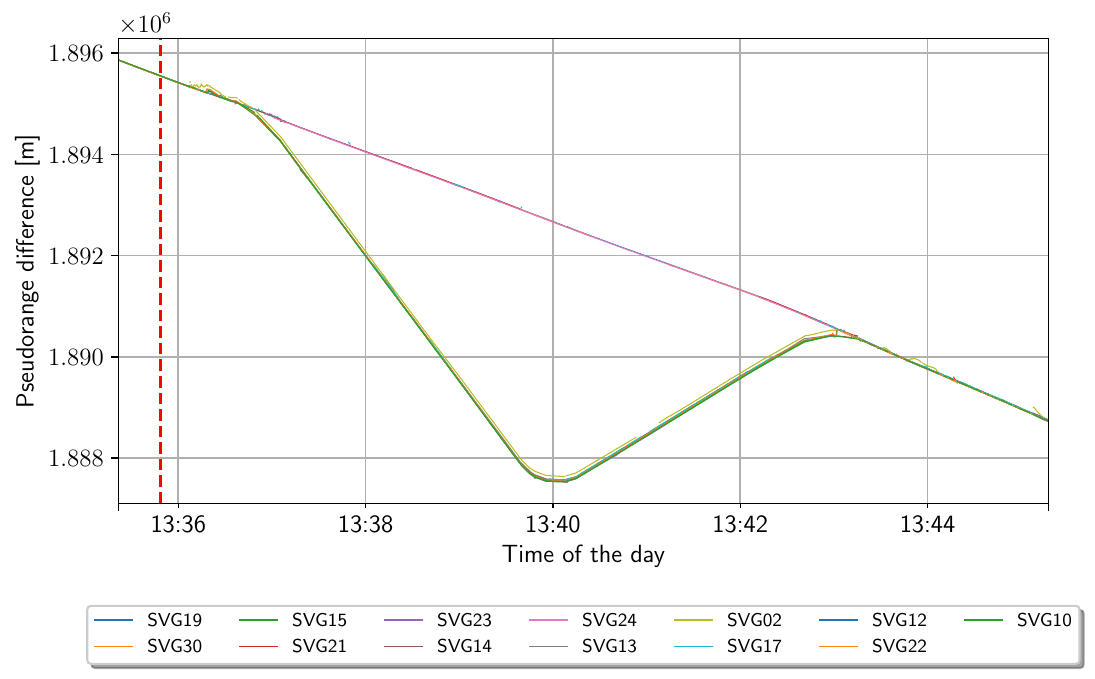}
    \caption{Pseudorange difference (reference vs target).}
    \label{fig:attacker-pr-strategy-rates}
\end{subfigure}%
\begin{subfigure}[b]{\columnwidth}
    \centering
    \includegraphics[width=\linewidth]{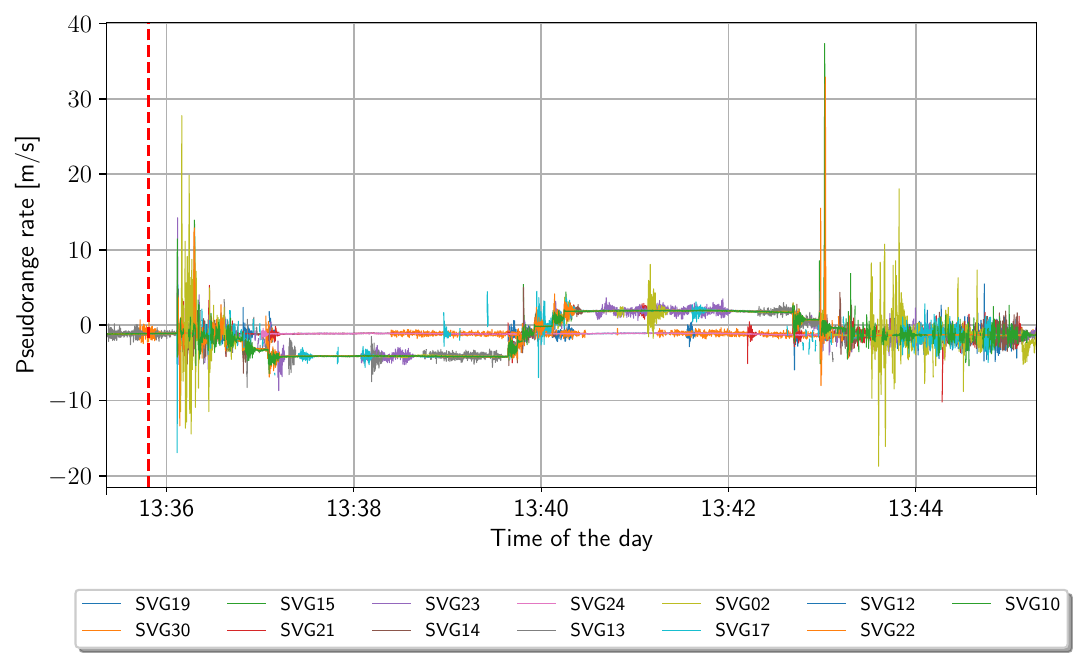}
    \caption{Pseudorange rate of change (reference vs victim).}
    \label{fig:attacker-pr-strategy-deltas}
\end{subfigure}
\caption{\textcolor{black}{Adversary control profile at the victim receiver and comparison with ground truth from a reference receiver, in terms of pseudorange difference (a) and pseudorange rate of change (b). The dashed line indicates the beginning of the attack.}}
    \label{fig:attacker-pr}
\end{figure*}

Control of the victim receiver is achieved by modifying the code phase and Doppler shifts of several or all satellites to achieve the pseudoranges matching the attacker's target PNT solution \cite{HumphreysAssessingSpoofer}. The modification of the pseudorange needs to be slow and progressive to avoid loss of lock at the receiver \cref{fig:attacker-pr-strategy}. There is a trade-off between attack aggressiveness (i.e. drift rate of the time solution at the GNSS receiver), attack objective, and access to the GNSS-enabled system that is unlikely known in advance and can favor the defense mechanism. A progressive stretch in the pseudoranges makes the victim receiver time solution lag behind the legitimate PNT, as the satellites are perceived to be further away from the victim receiver. Similarly, a decrease in pseudorange distance caused an acceleration of the time perception at the victim. \cref{fig:attacker-pr} show the mode of operation of an advanced attacker considered in this work: initialization of the simulation signals, the transmission of the attack signals, the start of the pseudorange ramp pull, an increase of the pseudorange ramp speed, stabilization of the attack and finalization of the adversarial objective.

\section{Results and Analysis}
\label{section:results-conclusion}

We organize the analysis in the following way: first, we discuss the profiling of the alternative time sources and the main limitations that arise regarding their use as time sources, both from an accuracy and security perspective. To achieve effective detection of misbehavior in GNSS-provided time, an analysis of the performance of the alternative time source is required. 
Second, we test the time-test countermeasure against a simplistic and an advanced attacker and discuss the effect such an attacker has when coherently tampering with the network link. Tests are performed with both NTP and NTS servers, when available. To the best of our knowledge, all servers provided by Netnod provide the same performance in NTP and NTS mode, with a minimal increase in computation caused by NTS at the client side (which, on most modern platforms, is negligible). Last, we discuss the limitations of the method and the role of the local clock in adaptive sampling for the remote time sources.

\subsection{Remote time providers classification and performance}
\begin{figure*}
    \centering
    \begin{subfigure}[t]{\columnwidth}
        \centering
        \includegraphics[width=\columnwidth]{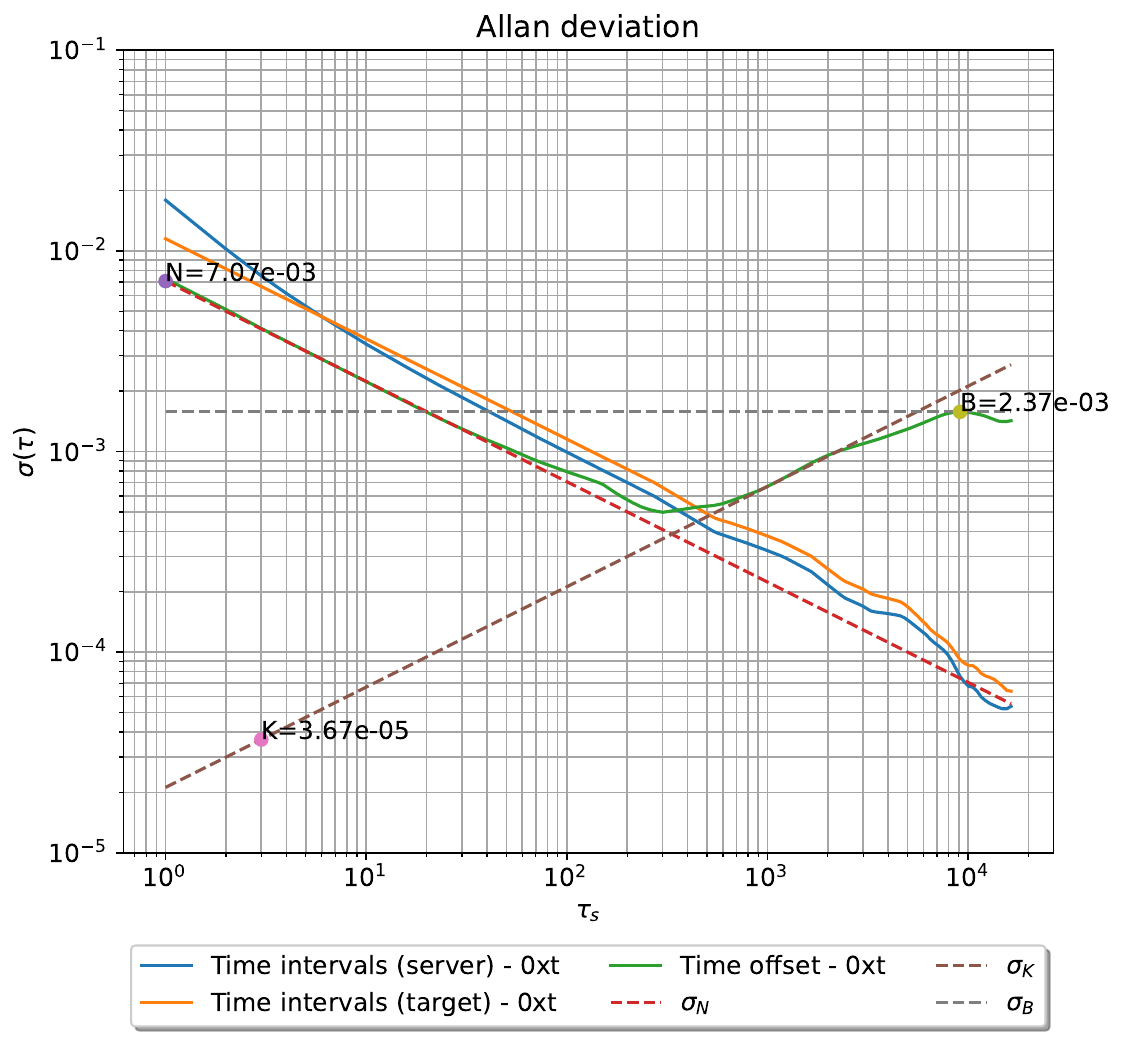}
        \caption{Roughtime server: time.0xt.ca}
        \label{fig:adev-0xt}
    \end{subfigure}%
    \begin{subfigure}[t]{\columnwidth}
        \centering
        \includegraphics[width=\columnwidth]{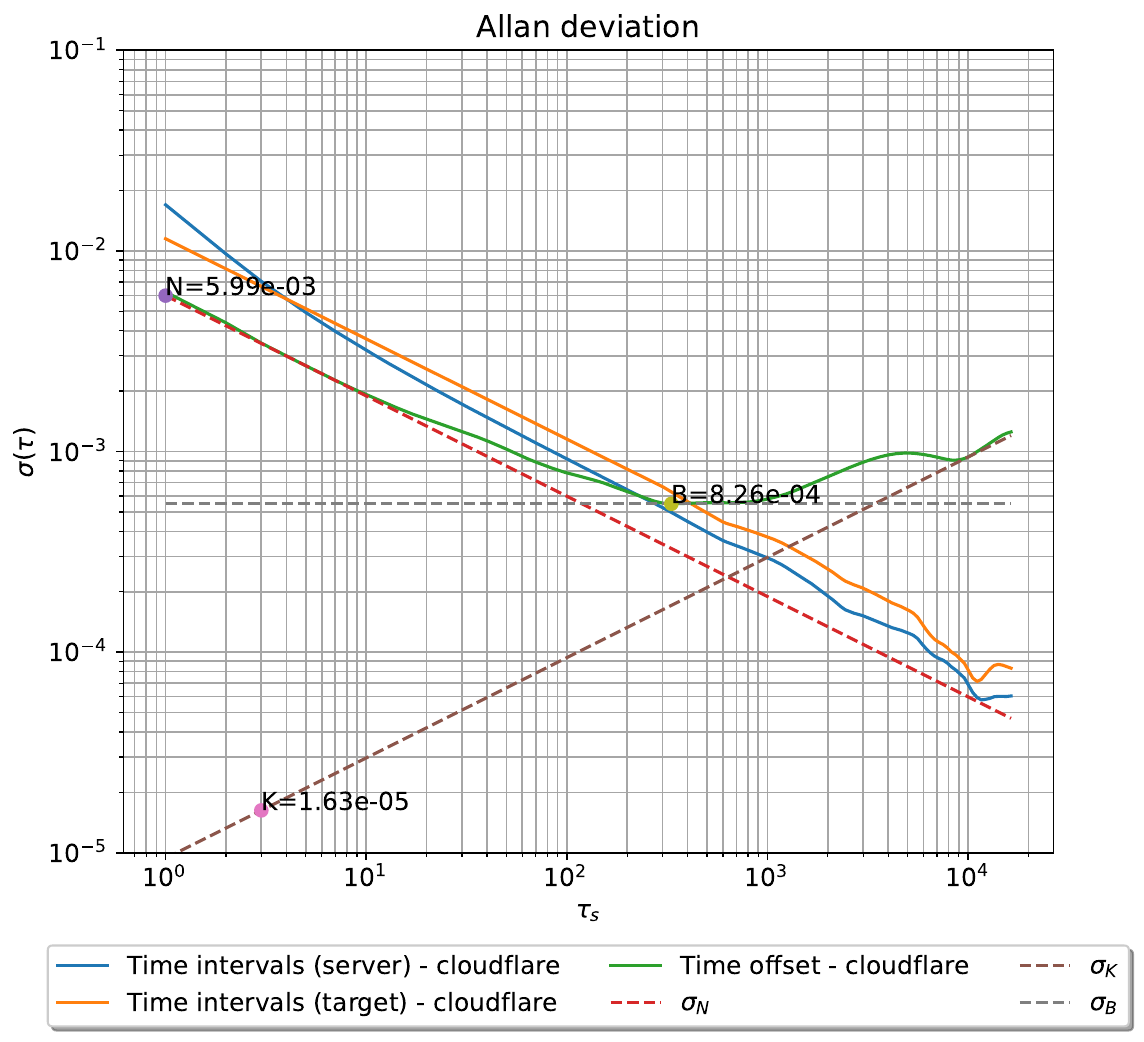}
        \caption{Roughtime server: roughtime.cloudflare.com}
        \label{fig:adev-cloudflare}
    \end{subfigure}
    \caption{Roughtime server stability (time.0xt.ca, left and roughtime.cloudflare.com, right) and local clock comparison (GNSS disciplined clock).}
    \label{fig:adev-rt}
\end{figure*}

As discussed in \cref{section:methodology}, different time providers available to the system can be classified based on availability, accuracy, and security level. These results will analyze how the performance of the remote (or local) time servers allows validation of the GNSS time solution: a thorough investigation on the quality of the public Roughtime and NTS infrastructures is worth a separate investigation. 
Notably, since the previous investigations \cite{spangheroMPPPLANS23}, the Roughtime ecosystem did not grow but instead, several servers are unavailable. This limits the possibility of testing different servers at different locations. We focus on two relevant time providers located in California (time.0xt.ca) and at Cloudflare (roughtime.cloudflare.com) which are the most reliable and available. \cref{fig:adev-0xt,fig:adev-cloudflare} show the Allan deviation (\cref{eq:allanvar}) for the server timestamp (reference), the local timestamp (target) and the inter-system offset. While Roughtime is not designed specifically for accuracy but mainly targets security and non-reliability of the time source, the servers exhibit better stability and accuracy than what is specified in the corresponding RFC (\cite{ietf-ntp-roughtime-07}). Although the servers perform similarly, \cref{fig:adev-0xt} shows comparatively worse stability due to the higher network latency, which becomes the dominant factor at longer integration times.

To be fully functional, Roughtime requires a healthy ecosystem of 3 or more servers to allow enchaining of the requests and cross-validation. At the moment, such infrastructure is not available or reliable (except for the two servers tested here), limiting the applicability of the system beyond a simple interval-based cross-check and majority agreement.

\begin{figure}
    \centering 
    \includegraphics[width=\columnwidth]{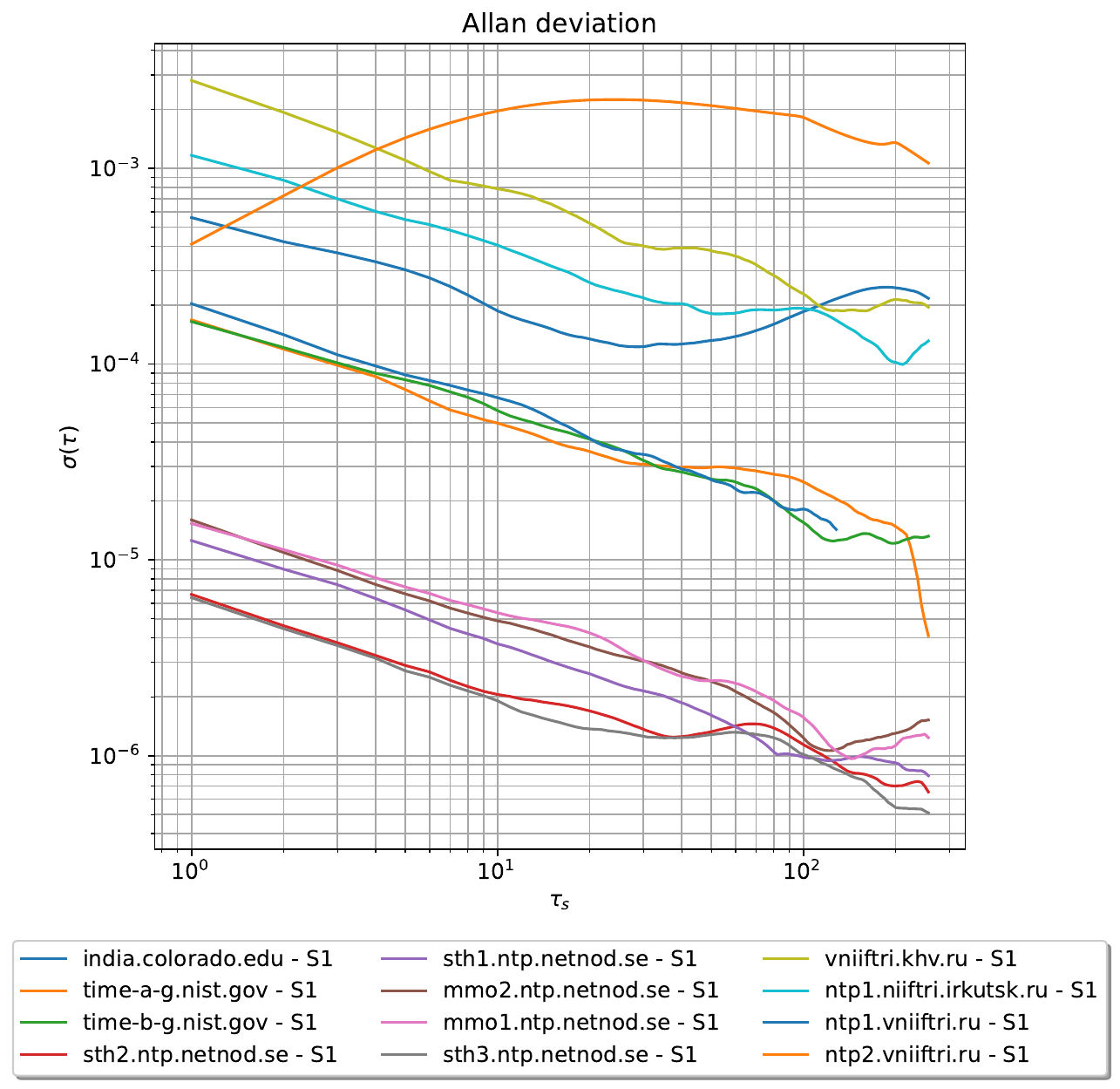}
    \caption{Allan deviation based on \cref{eq:allanvar}: servers tiers are defined by stratum and latency.}
    \label{fig:adev-ref-unspoofed}
\end{figure}

Similarly, the performance of NTP/NTS and NTS servers is also analyzed using \cref{eq:allanvar}. \cref{fig:adev-ref-unspoofed} shows the performance of geographically diverse servers over a publicly available internet connection. Specifically, we focus on five servers located in the same country as the measurement system which are provided by Netnod, and three servers provided by NIST. As in the Roughtime case, the network latency dominates the error, but in this case, the effect is more subtle. While the worst NTP/NTS server considered provides about three orders of magnitude short-term stability compared to the best Roughtime time reference, it is clear how NTS is less tolerant of network latency than Roughtime. In the perspective of validating the GNSS-provided time solution, NTP/NTS is well suited to perform online cross-checking of the time information and provides a higher quality reference. On the other hand, these results are dependent on the performance of the connection and the overall latency. Tests performed to simulate different levels of congestion show that increasing jitter causes the NTP/NTS time quality to progressively decrease at higher congestion rates.  

\label{subsec:performance-clustering}

\begin{figure}
    \includegraphics[width=\columnwidth]{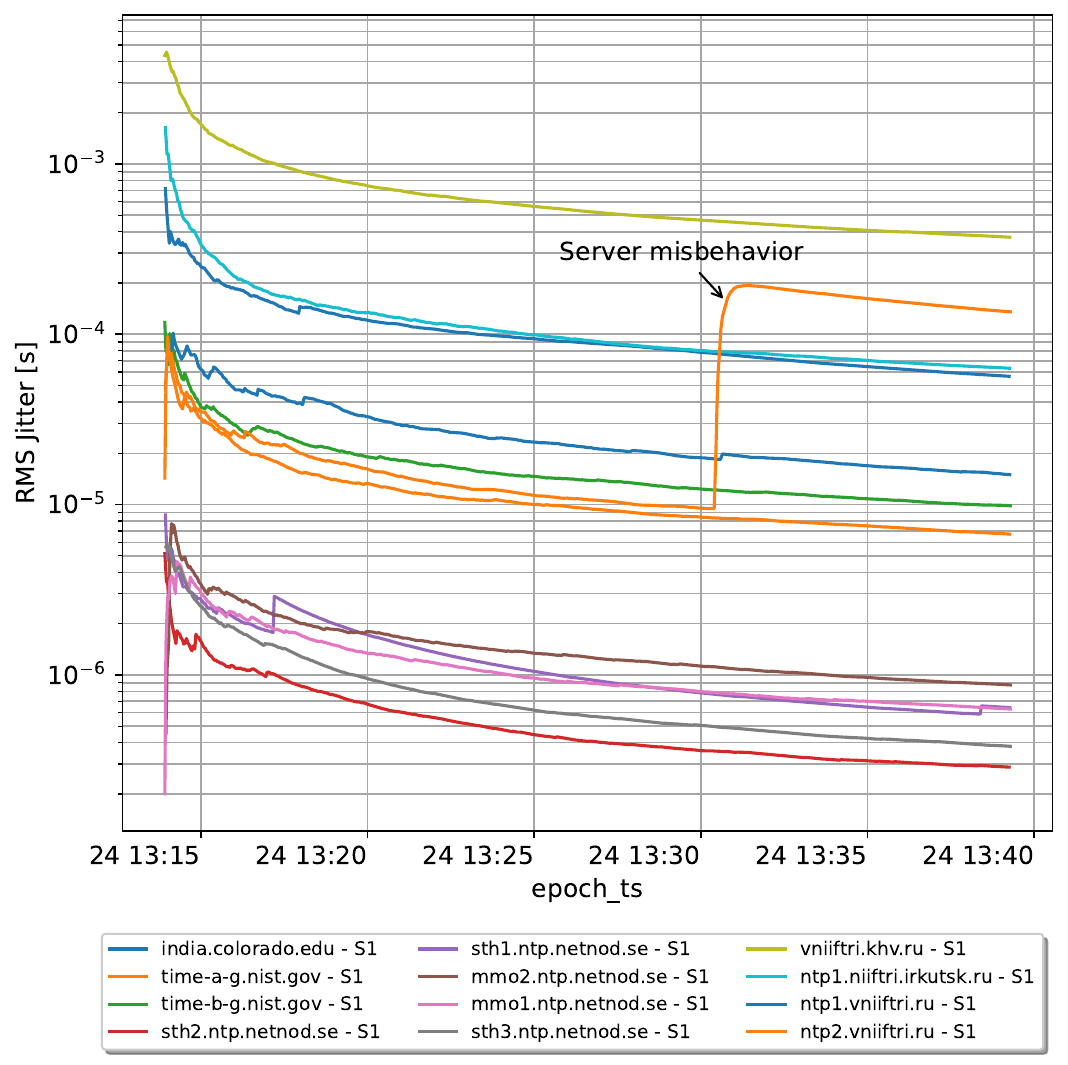}
    \caption{NTP/NTS reference servers RMS Jitter. One time reference misbehaves during the sampling.}
    \label{fig:jitter-rms-reference}
\end{figure}

Notably, in \cref{fig:adev-ref-unspoofed}, one of the time references is misbehaving (ntp2.vniiftri.ru) and can be easily rejected based on the fact that the provided true time is not within the expected region. This event is also shown in the RMS jitter measurement which is collected in real-time and shown in \cref{fig:jitter-rms-reference}.
Given the performance metrics extracted from the remote time reference the stratified approach presented in \cref{section:methodology} tackles two fundamental attacker behaviors: coarse asynchronous attacks and precisely aligned attacks.

\subsection{Performance under simplistic attack}
Due to the relatively low accuracy of Roughtime, such a validation method is preferable for the initial validation of the PNT solution (accordingly to \cref{sec:combining}). Contrary to NTP/NTS, where the client tracks a single source, Roughtime allows enchaining multiple requests to separate misbehaving servers. This allows the GNSS-enabled system to craft multiple requests to different servers at the same time using a nonce derived from the GNSS-provided timestamp. Furthermore, when estimating the coarse correctness of the GNSS-provided time, the roughness of the Roughtime server makes it robust to network latency for one-shot time checking. Combined with the robust cryptographic properties of Roughtime, this defeats practically any coarse adversary spoofing the time solution of the GNSS receiver, where coarse here is considered any attacker not capable of synchronization with the GNSS frame. 

\begin{figure}
    \includegraphics[width=\columnwidth]{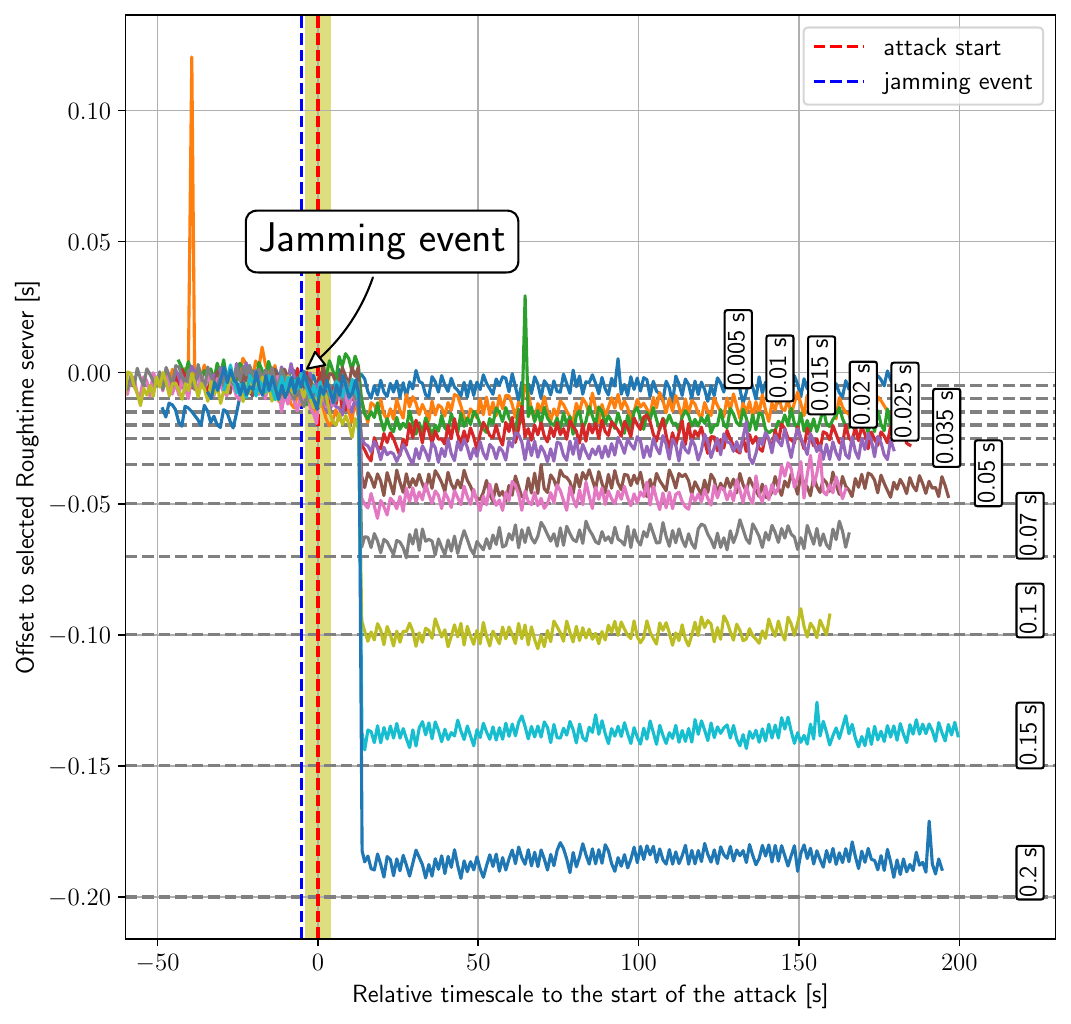}
    \caption{GNSS spoofing attack: increasing first fix time offset relative to the GNSS time (10 runs).}
    \label{fig:roughtime-checker-attack}
\end{figure}

\begin{figure}
    \includegraphics[width=\columnwidth]{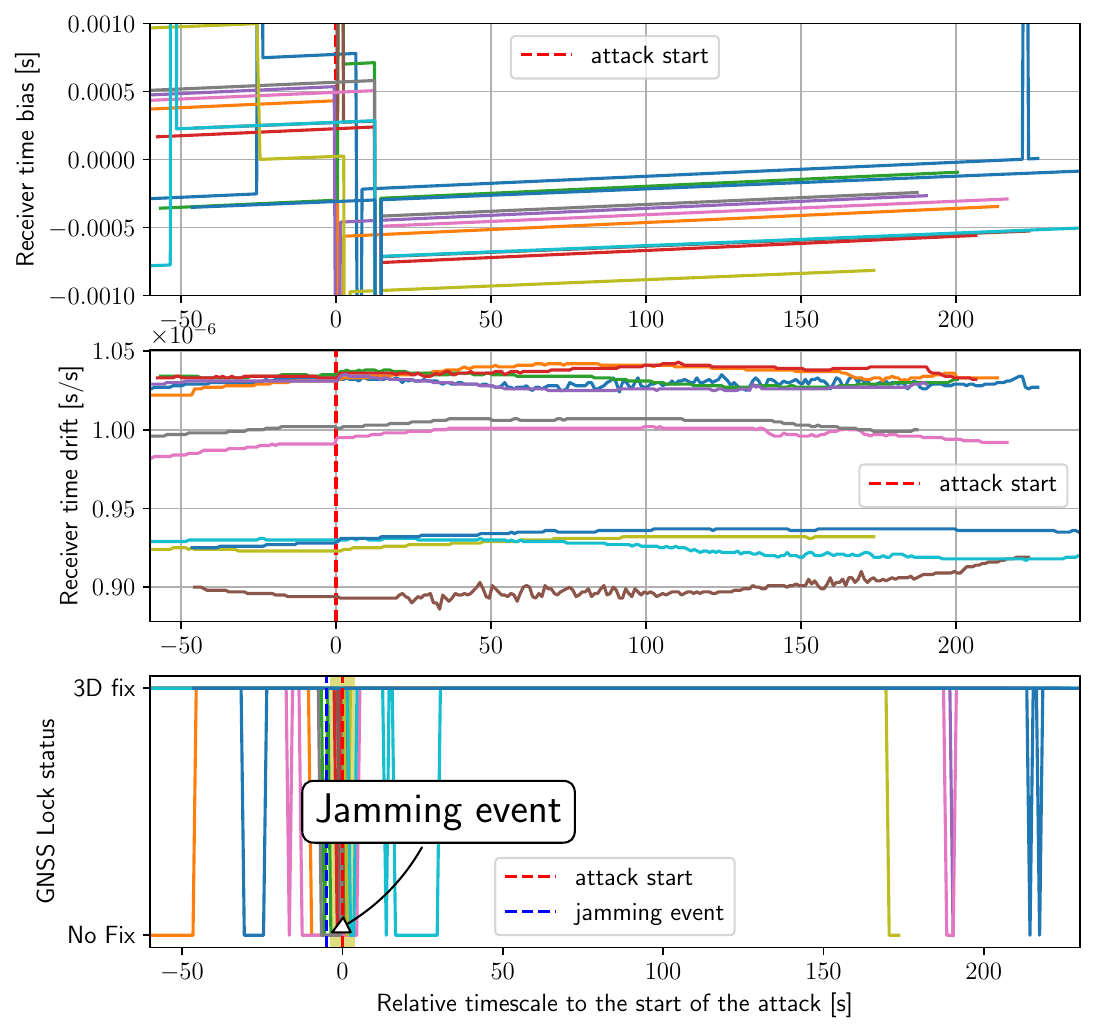}
    \caption{\textcolor{black}{GNSS spoofing attack: the GNSS receiver clock bias is stable even under attack (10 runs). The adversary delays the block transmission of the signals, but the internal clock offset (top) does not reflect the same change as the PNT solution. The clock drift (middle) is the same in the receiver before and after the attack start. Beyond the short initial jamming event, a valid PNT solution is provided throughout the entire attach (bottom).}}
    \label{fig:roughtime-checker-clkbias}
\end{figure}

\cref{fig:roughtime-checker-attack} shows the result of 10 attacks against a GNSS receiver, with increasingly higher offset between simulated signals and real signals GNSS frame. Notably, the Device under Test (DUT) is progressively more difficult to capture with decreasing synchronization of the simulated signals, requiring a short jamming burst to make the attack successful \cref{fig:roughtime-checker-clkbias}. Experimental evidence suggests that the required duration of the initial jamming phase is proportional to the stability of the GNSS receiver's local oscillator: the higher the quality of the GNSS receiver oscillator, the longer jamming is required. From \cref{fig:adev-0xt,fig:adev-cloudflare} even at short integration time, the detection threshold is set to tens of milliseconds and is bound by network latency to a few milliseconds at long integration times (several hundred seconds). 
\textcolor{black}{This is clear in \cref{fig:roughtime-checker-attack}, where attacks that maintain the initial synchronization at take-over below the Roughtime accuracy are not detected. This is because the detection threshold is set based on the measured accuracy of the Roughtime server (which proved to be better than the standard declared one). Generally, a detection threshold that is lower than the accuracy achievable at the reference time source will not be conclusive on the determination of adversarial manipulation of the GNSS time. In such conditions, e.g. $T_{off}=\mathrm{\SI{0.005}{\second}}$ in \cref{fig:roughtime-checker-attack}, the deviation caused by the attack is masked by the uncertainty of the reference.}

It is noteworthy that the initial bias measured with any Roughtime server is invariant for the same network delay, and can be validated based on initial cross-checking against any other Roughtime server. \textcolor{black}{While the clock bias of the GNSS receiver follows the one forced by the attacker after the GNSS receiver locks on the spoofed signals, as shown in \cref{fig:roughtime-checker-clkbias}, the drift is (in the short term) unmodified. This is consistent with the nature of the attack being only a time push and not a progressive modification of the time offset. Interestingly, the raw clock offset and bias at the GNSS receiver are unmodified by the attack, while the overall UTC time solution follows precisely the intentions of the adversary. Justification of such behavior is unknown without precise knowledge of the implemented PNT algorithm in the receiver but it shows how remote reference time checking is a valid augmentation towards providing assured PNT.}

Practically, if the GNSS-provided time fails this first check provided by coarse time testing, there is no point for the GNSS-enabled device to proceed with more advanced or precise testing solutions. Specifically, the GNSS-enabled system can reject GNSS time until the test keeps failing and rely on its onboard clock for timing.

\begin{figure}
    \includegraphics[width=\columnwidth]{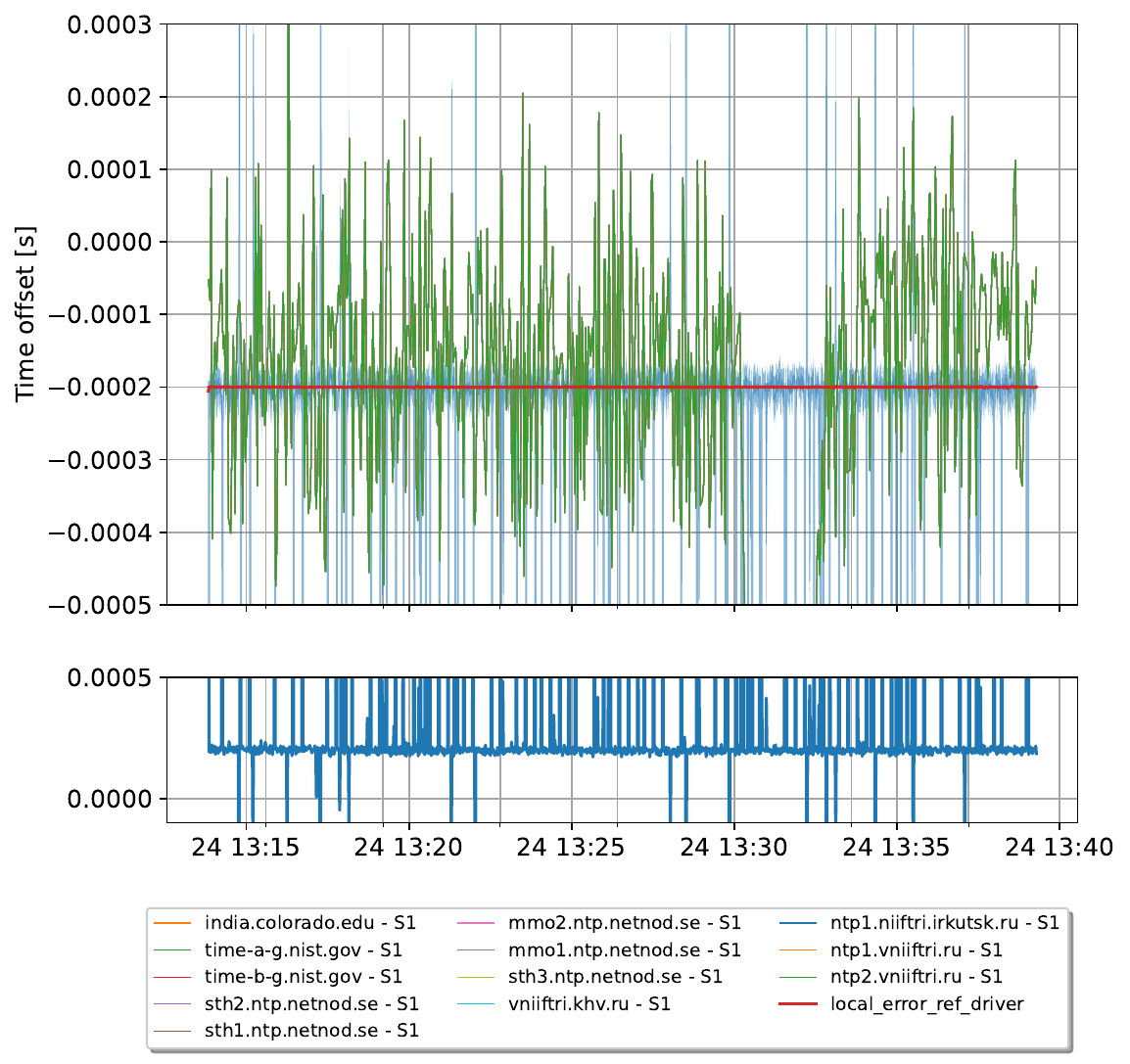}
    \caption{\textcolor{black}{NTP/NTS reference servers time check, where the consensus of time is based on the NTPv4 agreement scheme (top). Difference of consensus time against a local GNSS disciplined reference clock, in benign case (bottom).}}
    \label{fig:marzullo-consensus-reference}
\end{figure}

\begin{figure}
    \includegraphics[width=\columnwidth]{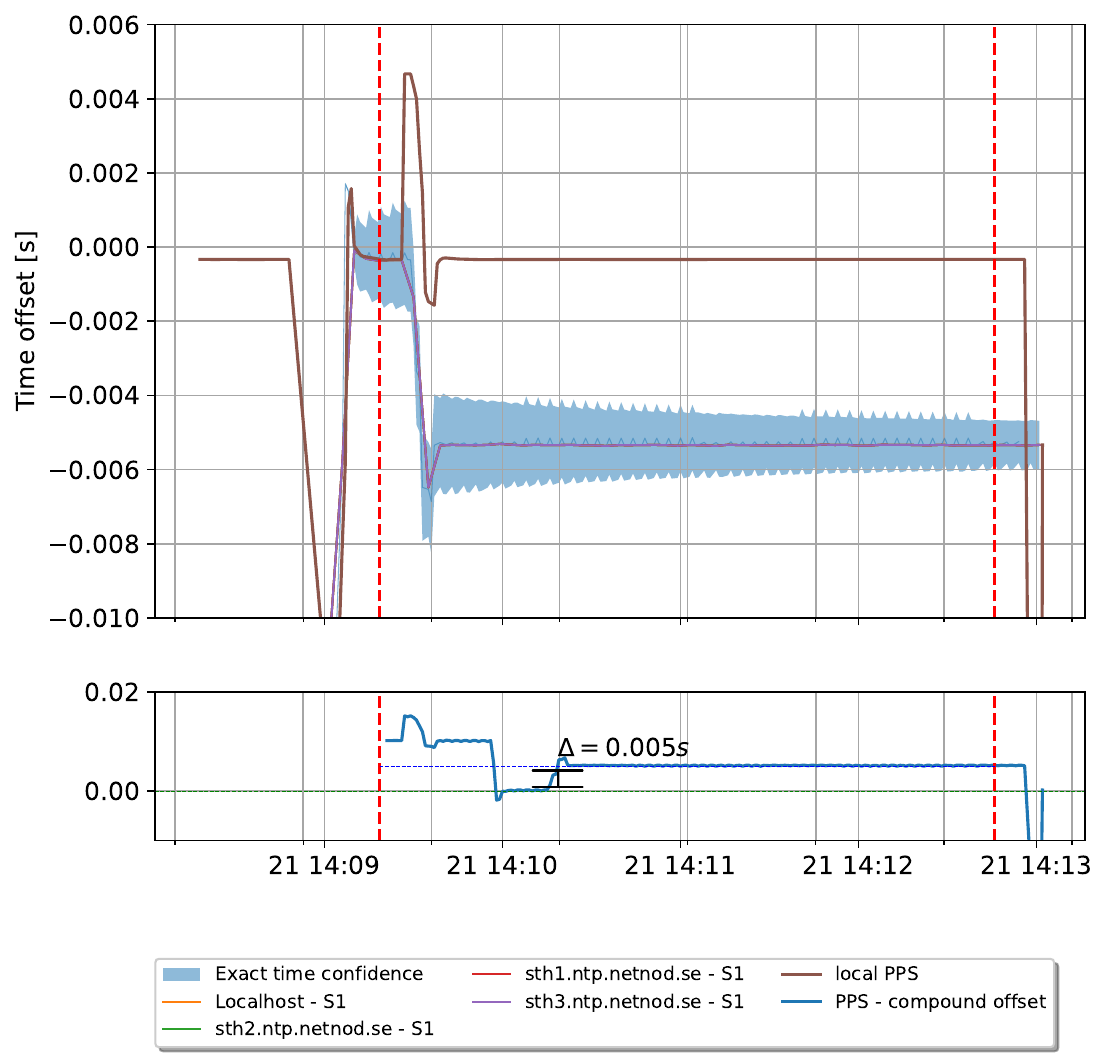}
    \caption{\textcolor{black}{NTP/NTS target servers time check: subtle time skip detection based on consensus (top). The difference between the mean reference server time and the GNSS receiver time solution shows the detection of a \SI{0.005}{\second} time push (bottom).}}
    \label{fig:marzullo-step-target}
\end{figure}

Similarly, NTP/NTS servers provide a check of the time solution but generally, convergence of Network Time requires multiple interactions. Nevertheless, if connectivity is present for a longer period, the NTP/NTS infrastructure can be used without source tracking to provide multiple redundant time references that are subject to consensus. In \cref{fig:marzullo-step-target}, the test with the lowest discontinuity ($T_{off}=\mathrm{\SI{0.005}{\second}}$) is repeated as a reference example to show how detection with NTP/NTP is also possible in case of abrupt variations in the GNSS-provided time solution.

Such an approach, based on \cref{section:marzullo}, is practical when several independent time sources are available, and provides a good estimate of the "exact time confidence" interval, as shown in \cref{fig:marzullo-consensus-reference}. Contrary to the Roughtime one-shot test, the estimate of the "correct" time interval improves with the refinement of the jitter measurement which extended to as many samples are available in the observation window. 

\subsection{Performance under synchronous attack}

\begin{figure}
    \includegraphics[width=\columnwidth]{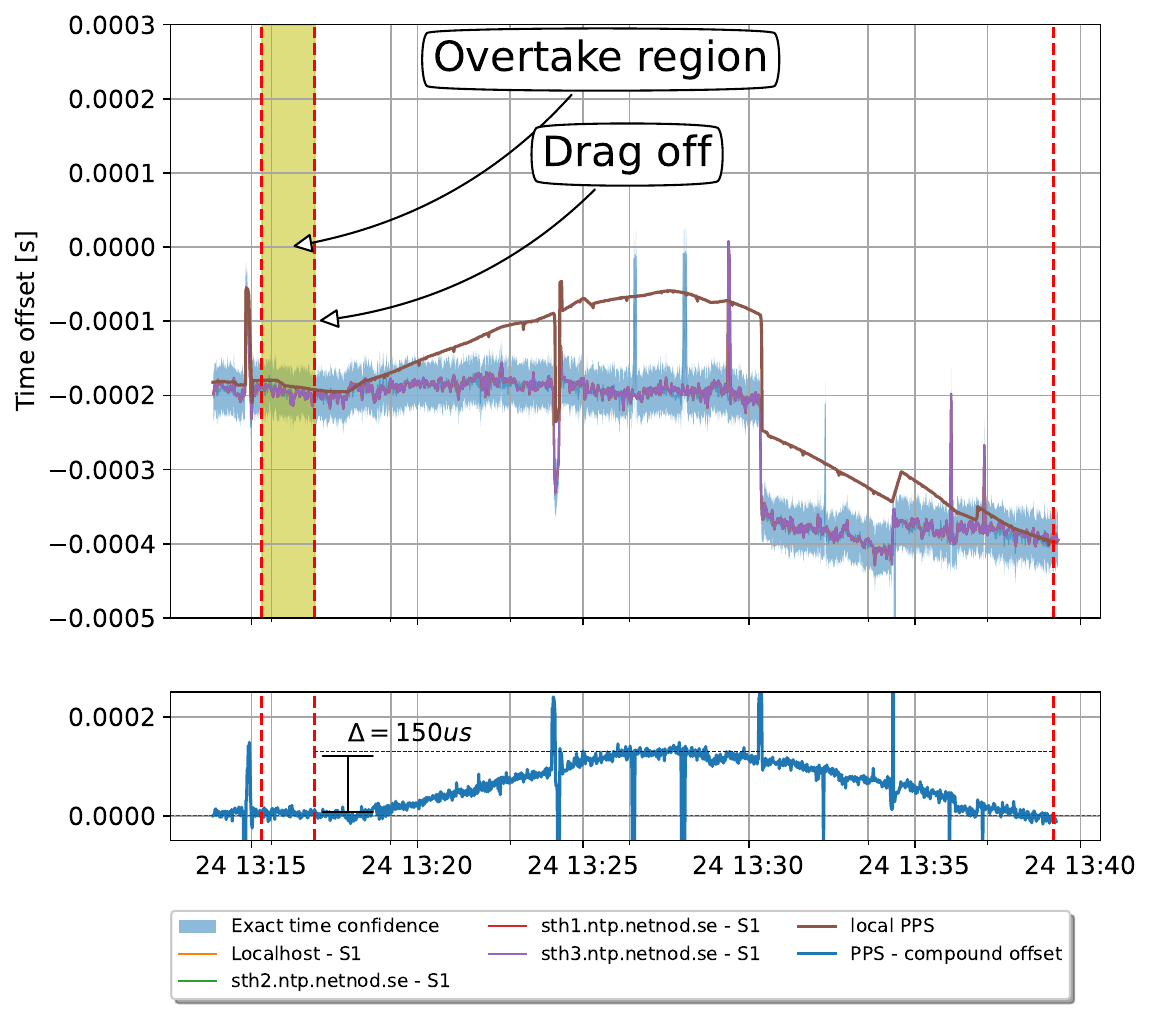}
    \caption{\textcolor{black}{NTP/NTS target servers time check: the GNSS-provided time under attack lies outside the interval of agreement of the reference servers (top). An attack forcing a \SI{150}{\micro\second} time shift is successfully detected (bottom).}}
    \label{fig:marzullo-consensus-target}
\end{figure}

Under attack, the GNSS receiver PPS/Time solution is manipulated by the adversary and will progressively drift away from the correct "true time" belief. This is shown in \cref{fig:marzullo-consensus-target}, where a \SI{150}{\micro\second} total adversarial deviation is first applied to the GNSS provided time and then progressively removed until the GNSS time is again synchronized with the "true time". Notably, the scenario tested here is quite advanced: the attacker not only drifts the time to its specific target but also guarantees that after the attack takes place there is no trace left. This requires the adversary to "roll back" its action and bring back the GPS timescale to the correct value. 

The total adversarial induced deviation is shown in the lower part of \cref{fig:marzullo-consensus-target}. The upper part shows how the attack is performed: the adversary first introduces the spoofing signals and allows the receiver to be captured by not forcing any deviation in the time solution. After roughly \SI{120}{\second} the adversary steadily drifts the time solution out of the "exact time confidence" interval. This is successfully detected by the compounding of network time providers, only when the total deviation crosses the boundary of the time confidence interval. To quantify how effective this approach is, it is important to recall how the time confidence interval is calculated. For each time server's time, the width of the single confidence interval depends on $\rho_d$, (\cref{section:marzullo}) which considers the total error at the source, the source jitter, and the network delay to the network time reference. Assuming that the time servers are globally synchronized with each other, the limiting factor becomes the network latency and the accuracy of the time reference, as all servers will agree on the same interval.

Even more subtle attacks focus on smaller modifications of the time solution and slower drag-out of the victim receiver from the legitimate GNSS-provided time. Under these hypotheses, an attacker can circumvent network-based time monitoring, it being either accurate or coarse. Additionally, a powerful attacker clogging or delaying the NTP/NTS messages consistently with the GNSS spoofing can produce enough latency at the victim system to make the GNSS attack unnoticed or deny high-quality network-based time references. In this case, a filter-based approach is more suitable as it continuously tracks the GNSS-provided time against the other time sources taking into account not only the absolute time of the reference but also the source time-varying characteristics.  The approach presented in \cref{eq:equation_test_single} for a local oscillator can be extended to an arbitrary number of reference clocks, either local or remote. Specifically, the parameters for tuning the Kalman filter can be extracted from the Allan deviation as the covariance expression is known for a generic oscillator \cite{Brown1991b}. Furthermore, the rate a which the Kalman estimator and tracking are executed can be tuned to comply with the resources available in the system and allow for missing measurements, which is important to consider in case of sudden loss of connection. In particular, the main drawbacks of executing the Kalman estimation sparsely are increased convergence time and detection latency, where the latter is defined as the time between the start of the attack and the first true positive of the hypothesis test based on \cref{eq:equation_test_single}, extended to all references available to the system. 

\begin{figure}
    \includegraphics[width=0.9\columnwidth]{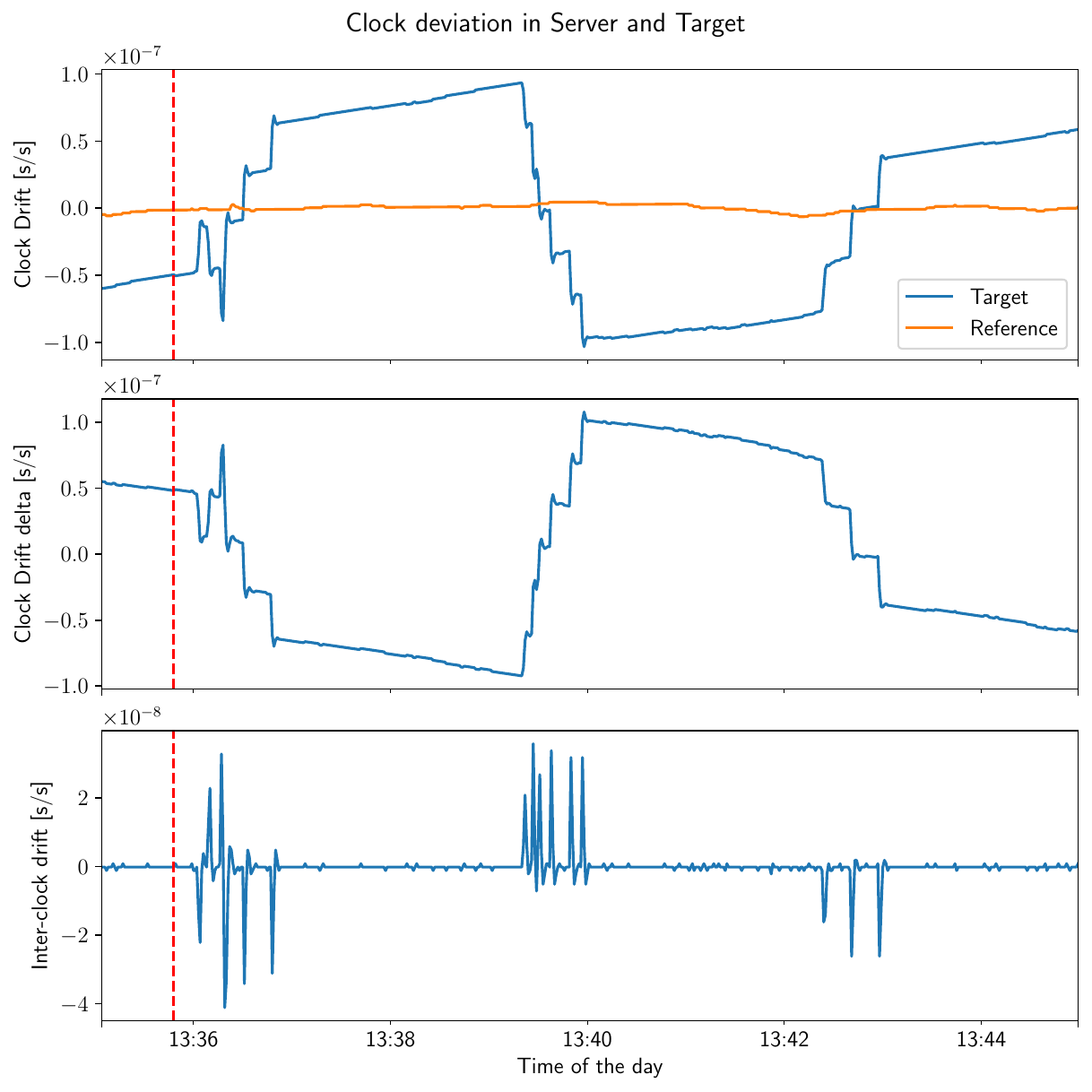}
    \caption{\textcolor{black}{Clock drift (top) in different GNSS-disciplined oscillators: the GNSS receiver under attack seamlessly follows the force deviation in the clock parameters. Deviation and attacker's pull rate is measured in comparison with a local hardware clock (middle, bottom).}}
    \label{fig:clock-drift-inter-rx}
\end{figure}

On the other hand, continuous tracking with a stable local reference source guarantees that the GNSS-receiver clock behaves with the specified parameters. This is seen in \cref{fig:clock-drift-inter-rx}, where the GNSS-disciplined oscillator is measured against another reference oscillator. At the start of the attack, the adversary mimics the current clock drift and progressively pulls the GNSS receiver away from the correct solution. \textcolor{black}{The clock drift in the victim follows the one forced by the attacker \cref{fig:clock-drift-inter-rx}, top). The difference against the time reference is used for detection (\cref{fig:clock-drift-inter-rx}, middle). The inter-clock drift provides information in regards to the attack's pull rate (\cref{fig:clock-drift-inter-rx}, bottom).}

\begin{figure}
    \centering 
    \includegraphics[width=0.9\columnwidth]{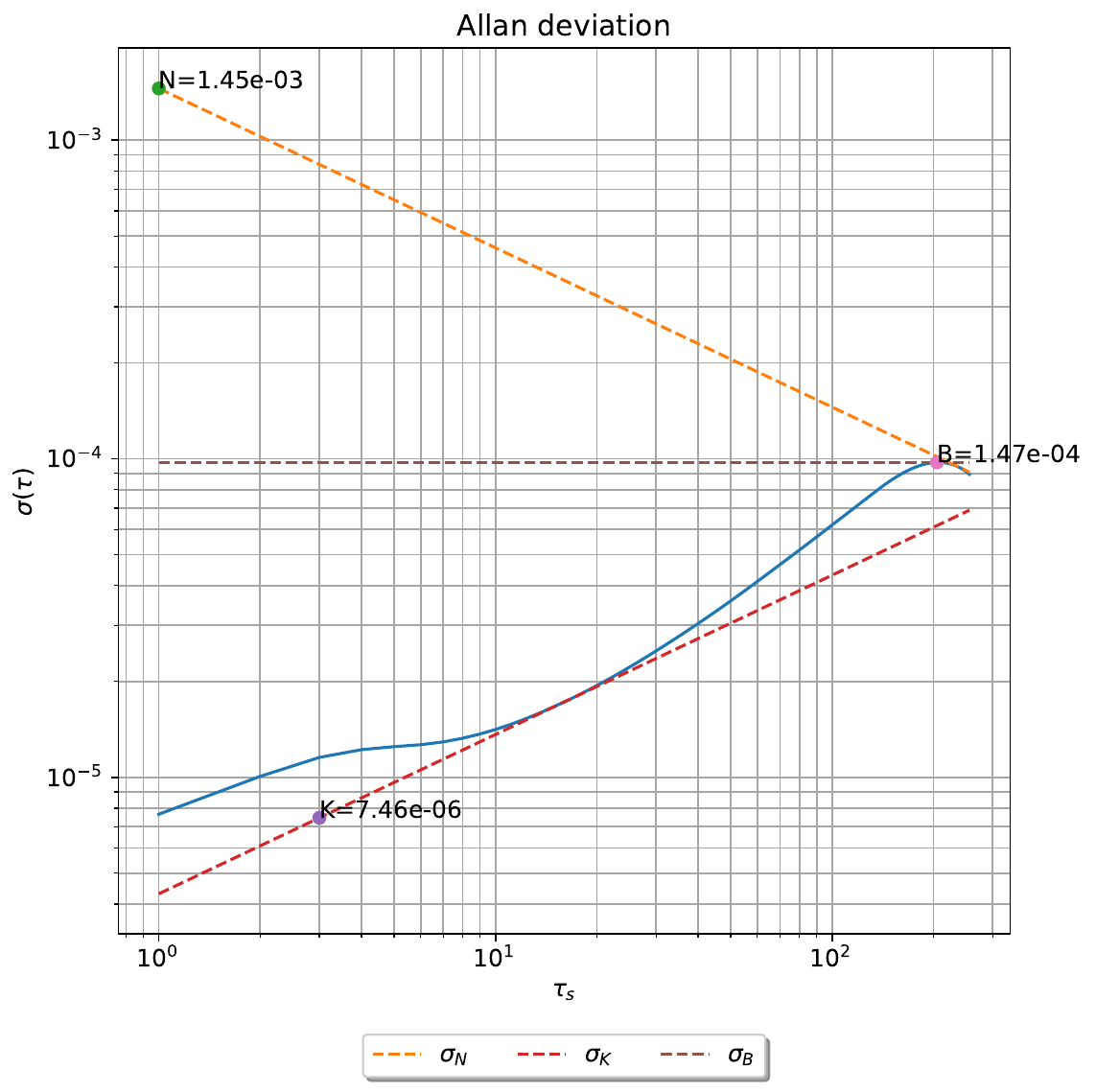}
    \caption{Allan deviation based on \cref{eq:allanvar} for GNSS PPS source, target receiver under GNSS spoofing attack.}
    \label{fig:adev-tgt-spoofed-pps}
\end{figure}

Additionally, a local time reference cannot be tampered with by an attacker that does not have access to the victim device, it is immune to adversarial network delay and can be used to provide holdover information during network outage. \cref{fig:adev-tgt-spoofed-pps} shows the Allan deviation of the ensemble between the GNSS-receiver disciplined oscillator and a local clock, under the same spoofing conditions as in \cref{fig:clock-drift-inter-rx}. Compared to the benign case in \cref{fig:adev-ref-unspoofed}, where the GNSS receiver is not under the control of the attacker, the Allan deviation difference is clear.


\subsection{Latency attacks on the network link}

Additionally, the attacker can control the link latency by setting up a clogging attack on the network connection between the Internet and the target device. Experimental evidence shows that it is not necessary to directly clog the receiving device: inducing latency in the switch the device uses to access the network is sufficient to cause a significant degradation in the latency. When cryptographic countermeasures are in place, a clogging attack is the simplest and most effective way of decreasing the accuracy and availability of a remote time source. The test is conducted over a commercial 1Gbps ethernet connection, with progressively increasing traffic up to the switch's capacity. During this test, no public server was affected, to avoid service disruption, and all testing was done on the client side. 

While the latency in the communication increases, the Roughtime server timestamps do not change in accuracy. This is achieved by the design in Roughtime, which only certifies the time at the server, without considering the link latency. Additionally, the increased latency only causes the server to drop requests that time out. Even in a challenging network environment, the device is still able to query the Roughtime server despite the almost saturated link.

\begin{figure}
    \includegraphics[width=\columnwidth]{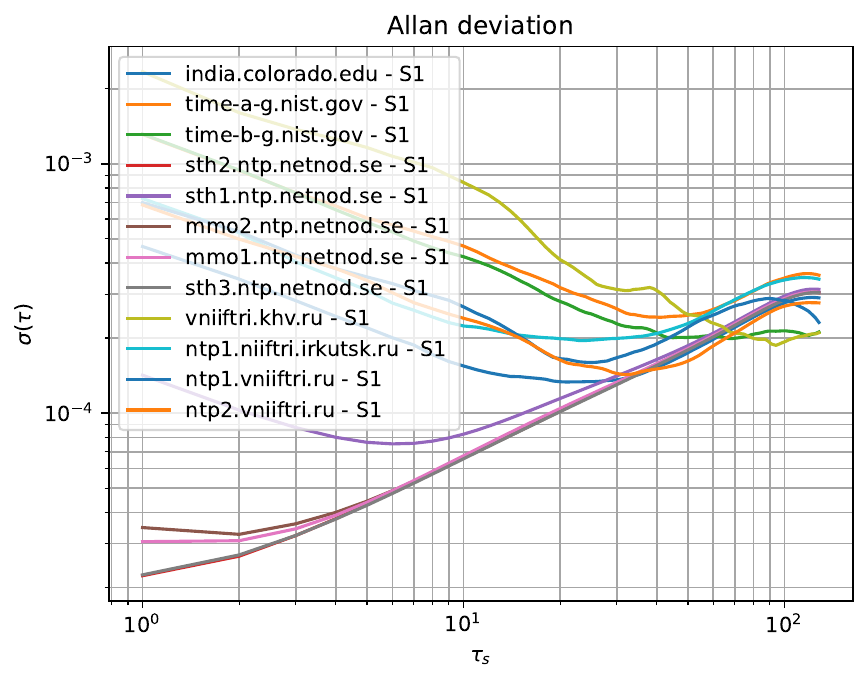}
    \caption{NTP/NTS servers time check under heavy load. The local traffic-related latency heavily reduces accuracy, aiding a multi-surfaced attacker.}
    \label{fig:ntp-clogged}
\end{figure}

In comparison, \cref{fig:ntp-clogged} shows the Allan variance measurements for the same set of NTP/NTS severs as in \cref{fig:adev-ref-unspoofed}, but under a clogging attack. In this case, there is a considerable change in the accuracy of the externally provided time: the long-term accuracy decreases constantly, as opposed to the non-clogged case. 

\subsection{Adaptive sampling of remote time sources}
\label{subsection:adaptive-sampling}

At this point, it is interesting to consider if it is necessary to test every single PNT update from the GNSS receiver, or if it would be better to sample-test some of the updates. External network-based time references often limit the frequency at which a client can perform the synchronization protocol, to guarantee fairness among multiple clients connected to the same server. To the best of our knowledge, most commercial providers limit the minimum re-synchronization interval to \SI{1}{\second}, dropping requests that happen more often than that, forcing the client to back-off. Generally, commercial GNSS receivers for civilian use provide a PNT update with a frequency of \SI{1}{\hertz}, which is compatible with the NTP/NTS server rate. Practically, if the PNT updates happen more frequently than the server's allowed re-synchronization rate, the device can still rely on its onboard clock to validate every update, and on the external time source to validate the updates at a slower rate, by decimation.  

If the adversary modifies the PNT solution by applying slow drifts (e.g. $<$\SI{10}{\nano\second/\second}) the modification is slow enough to avoid detection from any of the presented methods. In such a situation the GNSS-enabled device, given that it is provided with a clock that is stable enough or it has access to a quality remote reference, instead of estimating the error of the current estimate can run a two-point interpolation approach deriving the error estimation parameters from the Kalman filter. Intuitively this approach is simple: the receiver saves a sample and checks if the time interval at the GNSS receiver provided timebase and at the remote timebase match after a pre-defined sampling period. This method allows leveraging long- and short-term stable reference clocks independently. This is equivalent to sampling with memory: the GNSS attack action needs to be smooth and constant, or it would be detected by the presented methods. The device can test samples at different intervals (i.e., two samples \SI{1}{\second} apart, two samples \SI{10}{\second} apart, two samples \SI{100}{\second} apart) and estimate the interval of time elapsed at the GNSS receiver and the alternative time providers. This allows to amplify the effects of the attacker, making the proposed countermeasure more sensitive.
A score is defined based on the number of successful validations of the GNSS provided time: at each successful test, the next polling interval is extended by the same value. Practically, this causes a linear decrease in the remote sources' polling frequency, up to a pre-defined threshold. Progressively, the solution is monitored more sparsely, effectively saving bandwidth and processing power. If at any given sample, the validation test fails, the adaptive sampling rate is scaled back to the minimum interval, more closely monitoring the time solution. 

\subsection{Security considerations and overheads}
\label{subsection:security-considerations}

Based on \cref{section:fusion-tradeoff}, the GNSS-enabled system can always trade accuracy for time assurance, where the latter is defined as the level of certainty the system has regarding both the accuracy and the trustworthiness of the time offset estimate.
This requires careful analysis of the security level of the remote time sources. Excluding the local reference oscillator, which is integral to the GNSS-enabled system and can be assumed trusted, the network-based time references are subject to various levels of adversarial control. Cryptographical enhancements allow network-based time sources to provide secure and assured time even if the network link is potentially adversarial.  Roughtime's use of asymmetric cryptography introduces a significantly higher computational cost compared to other methods (secureNTP, NTS), but the overall latency is strongly dominated by the network component, while the cryptographic information validation only accounts for a small amount, in particular on modern mobile CPUs. 

Performance-wise, the main concern is regarding the cryptographic overhead introduced by the new secure time distribution system. Overall, it is manageable even by small SWAP platforms (e.g., ours presented in \cref{section:implementation})
If a GNSS-enabled system is heavily constrained and is incapable of continuously running a cryptographically secure protocol, it can adapt the sampling rate of the remote time sources based on the Allan deviation estimation (\cref{section:results-conclusion}, \cref{subsection:adaptive-sampling}). Stable time reference systems offer similar performances at longer sampling rates, which overall reduce the average cryptographic overhead. 

An advanced adversary could potentially control a GNSS-enabled device if operating a colluding time reference server without resorting to advanced spoofing. Practically, the attacker establishes a time server within the ecosystem that reports an authenticated time consistent with the attacker's modification of the GNSS receiver time \cref{section:sys-adversary-model}. While advanced, this mode of operation is easily defeated by either a healthy ecosystem where multiple time servers are available or by further checking. Network-based time references lacking security enhancements are still beneficial for validating the GNSS-provided time solution, especially when combined with a few trusted sources or when network access is cryptographically protected.  

At the current level of implementation, only a handful of servers comply with NTS or Roughtime, making it difficult to provide ubiquitous secure time distribution. The unprotected implementation of NTP is still a valid choice to obtain many, geographically convenient time providers especially when their time offset can be validated against a single trusted time provider.

\section{Conclusion and Future Work}
\label{section:conclusion}
In this work we improve the method from \cite{spangheroMPPPLANS23} and provide a comprehensive evaluation detecting adversarial manipulation of the GNSS time, based on onboard and external/remote time source (accessible through the platform that encompasses the GNSS receiver). We examine both the individual performance and combination of three existing components for GNSS time-based validation in different scenarios. Our approach strongly limits the possibility of undetected (asynchronous or synchronous) simulation-based attacks targeting the GNSS receiver and, in addition, the network components. Attacks were successfully detected in all tests performed, either when the time solution was synchronously spoofed by the adversary or forced with a step change even in adverse network conditions. 

An evaluation of the security properties and overheads of secure time transfer and digitally signed remote time references shows that modern platforms can easily rely on secure network-provided time to validate the GNSS solution. Multiple combinations are possible, by leveraging different secure remote time providers. When connectivity to remote time references is not available or attackers controlling the network communication are present, onboard reference sources allow continuous monitoring of the time part of the PNT solution. 

The value of detecting attacks based on consistency with external time source is broadly applicable, not only when the adversary targets the GNSS receiver time. To the best of the authors' knowledge, a trauma on the time offset is often present at the take-over stage, even if the latter is smooth. On the other hand, adversaries capable of modifying the PN part of the GNSS solution, without modifying the time solution cannot be detected by countermeasures based on time validation. This is true for the method presented here or any other time-based countermeasure. In conclusion, this method can easily be deployed to protect existing receivers from various adversaries, without modifications to its structure.

\bibliographystyle{ieeetran}
\bibliography{refactorbib}

\begin{IEEEbiography}[{\includegraphics[width=1in,height=1.25in,clip,keepaspectratio]{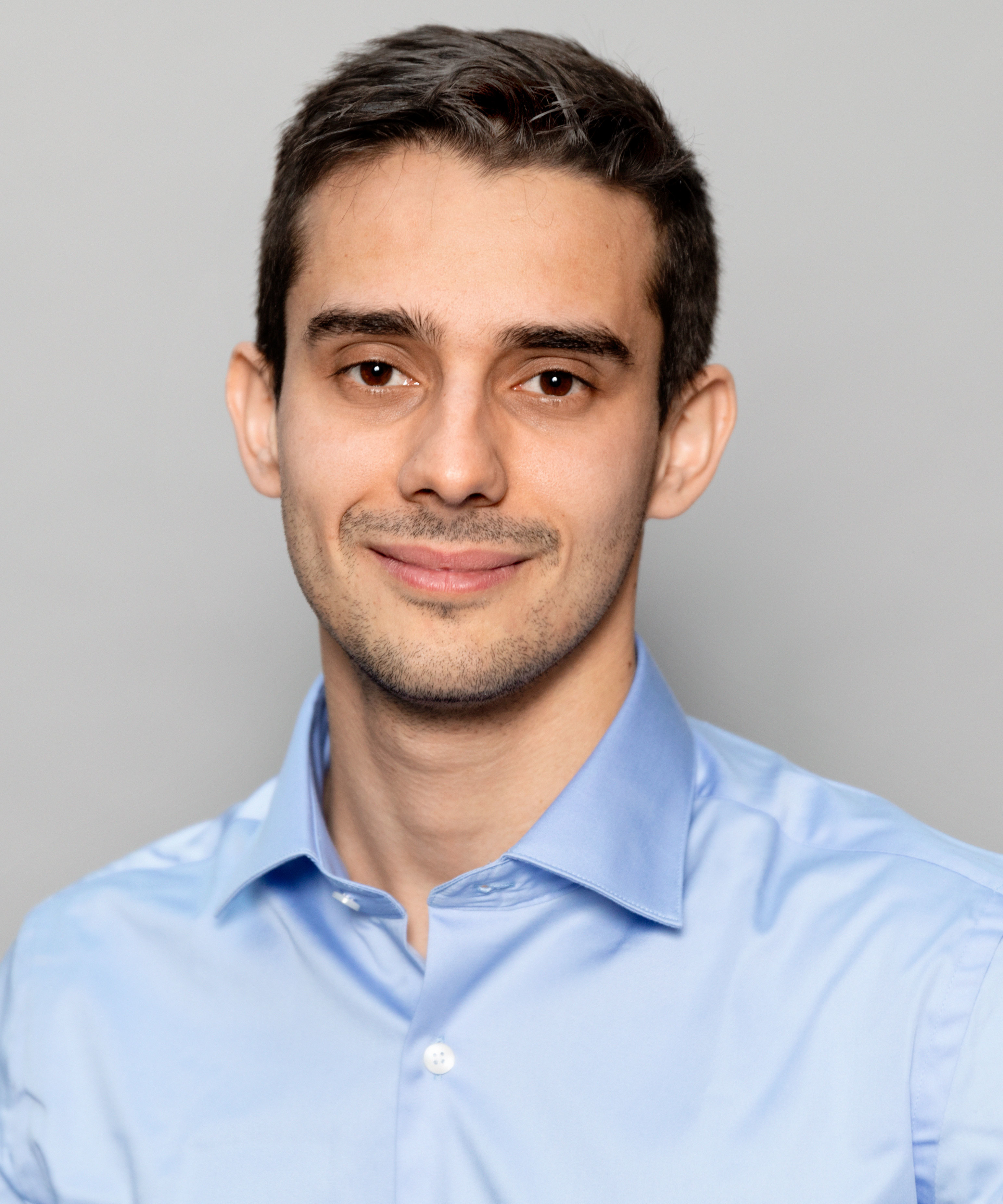}}]{Marco Spanghero} received his B.S. from Politecnico of Milano and an MSc degree from KTH Royal Institute of Technology, Stockholm, Sweden. He is currently a Ph.D. candidate with the Networked Systems Security (NSS) group at KTH, Stockholm, Sweden, and associate with the WASP program from the Knut and Alice Wallenberg Foundation. 
\end{IEEEbiography} 
\vskip -2\baselineskip plus -1fil
\begin{IEEEbiography}[{\includegraphics[width=1in,height=1.25in,clip,keepaspectratio]{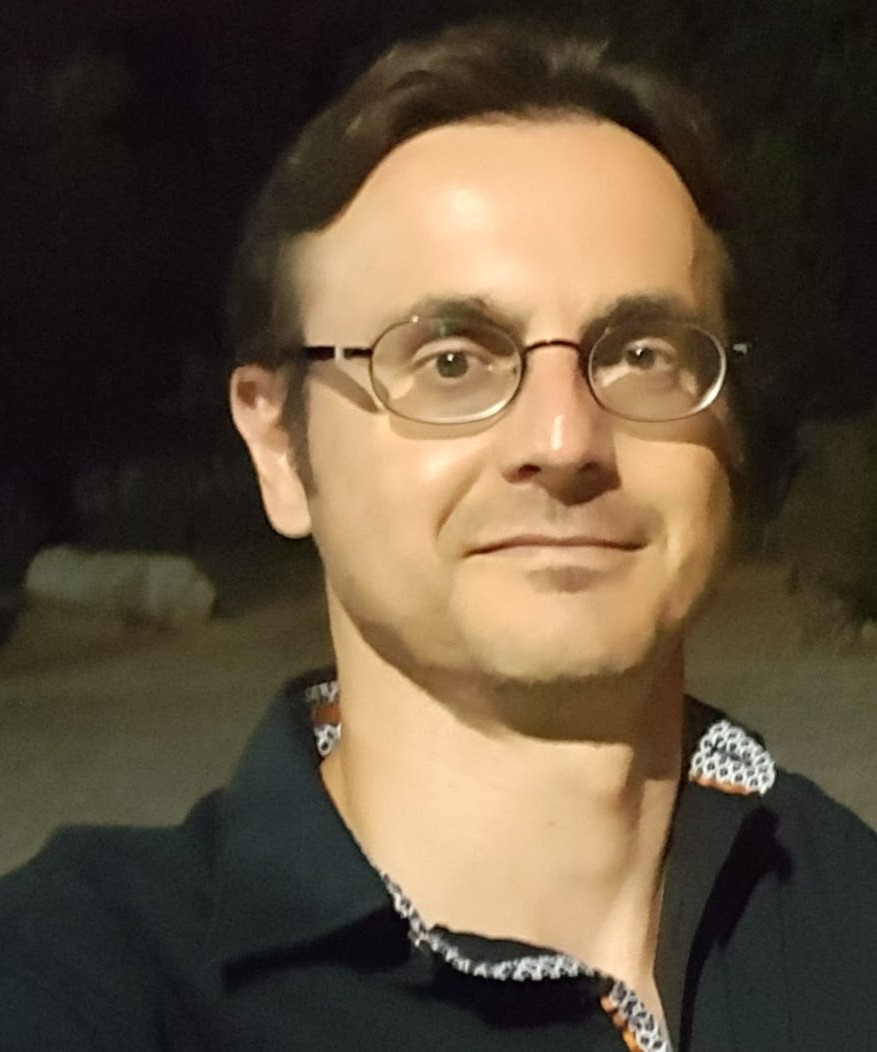}}]{Panos Papadimitratos}
	(Fellow, IEEE) earned his Ph.D. degree from Cornell University, Ithaca, NY, USA. At KTH, Stockholm, Sweden, he leads the Networked Systems Security (NSS) group and he is a member of the Steering Committee of the Security Link Center. He serves or served as a member of the ACM WiSec and CANS conference steering committees and the PETS Editorial and Advisory Boards; Program Chair for the ACM WiSec’16, TRUST’16, and CANS’18 conferences; General Chair for the ACM WISec’18, PETS’19, and IEEE EuroS\&P’19 conferences; Associate Editor of the IEEE TMC, IEEE/ACM ToN and IET IFS journals, and Chair of the Caspar Bowden PET Award. Panos is a Fellow of the Young Academy of Europe, a Knut and Alice Wallenberg Academy Fellow, and an ACM Distinguished Member. NSS webpage: \url{https://www.eecs.kth.se/nss}.
\end{IEEEbiography} 
\end{document}